\documentclass[11pt]{article}
\usepackage{authblk}
\usepackage{natbib}
\usepackage{amsmath, amsfonts,mathrsfs}
\usepackage{color}
\usepackage{rotating}
\usepackage{JASA_manu1}
\usepackage{lscape}
\usepackage{graphicx,multirow}
\usepackage{bm,url}
\usepackage{booktabs}
\usepackage{caption}
\usepackage{subfig}

\newtheorem{lemma}{\noindent\mbox{Lemma}}
\newtheorem{theorem}{\noindent\mbox{Theorem}}
\newtheorem{corollary}{\noindent\mbox{Corollary}}
\newtheorem{proposition}{\noindent\mbox{Proposition}}
\newtheorem{condition}{\noindent\mbox{M}}

\DeclareMathOperator*{\argmax}{arg\,max}
\def\be{\begin{equation}}
\def\ee{\end{equation}}
\def\bea{\begin{eqnarray}}
\def\eea{\end{eqnarray}}
\def\bd{\begin{displaymath}}
\def\ed{\end{displaymath}}
\def\bda{\begin{eqnarray*}}
\def\eda{\end{eqnarray*}}
\def\bsm{\begin{small}}
\def\esm{\end{small}}

\def\ha1{\hat \beta_1}

\def\bb0{\delta_\beta}

\def\bsc{\begin{scriptsize}}
\def\esc{\end{scriptsize}}

\begin{document}

\title{Change Point Detection in the Mean of High-Dimensional Time Series Data under Dependence}
\author[1]{Jun Li}
\author[2]{Minya Xu}
\author[3]{Ping-Shou Zhong}
\author[1]{Lingjun Li}
\affil[1]{Department of Mathematical Sciences, Kent State University, Kent, OH 44242}
\affil[2]{Guanghua School of Management, Peking University, Beijing 100871, China}
\affil[3]{Department of Statistics and Probability, Michigan State University, East Lansing, MI 48824}
\date{}
\renewcommand\Affilfont{\itshape\normalsize}
\maketitle

\begin{center}
\textbf{Abstract}
\end{center}

High-dimensional time series are characterized by a large number of measurements and complex dependence, and often involve abrupt change points. We propose a new procedure to detect change points in the mean of high-dimensional time series data.  
The proposed procedure incorporates spatial and temporal dependence of data and is able to test and estimate the change point occurred on the boundary of time series. 
We study its asymptotic properties under mild conditions. Simulation studies demonstrate its robust performance through the comparison with other existing methods.  
Our procedure is applied to an fMRI dataset.  

 \vspace*{.1in}


\setcounter{section}{1} \setcounter{equation}{0}
\section*{\large 1. \bf Introduction}

Many dynamic processes involve abrupt changes and change point analysis is to identify the locations of change points in time series data. There exists abundant research on change point analysis for univariate time series data. Examples include Sen and Srivastava (1975), Incl\'an and Tiao(1994), Chen and Gupta (1997), Kokoszka and Leipus (2000), Lavielle and Moulines (2000), Ombao et al. (2001), Davis et al. (2006), Davis et al. (2008), and Shao and Zhang (2010). Change point analysis for classical  multivariate time series data has also been extensively studied.
Examples include Srivastava and Worsley (1986), James et al. (1992), Desobry et al. (2005), Harchaoui et al. (2009), Zhang et al. (2010), Siegmund et al. (2011) and Matteson and James (2014).

With explosive development of high-throughput technologies, high-dimensional time series data are commonly observed in many fields including medical, {environmental}, financial, engineering and geographical studies. Change point analysis for high-dimensional data has received a lot of attention in recent years. 
For instance, 
Bai (2010) 
considered estimating the location of a change point in high-dimensional panel data under the assumption that the change has occurred {\it a priori}. 
Chen and Zhang (2015) proposed a graph-based approach to test and estimate change points under the assumption that a sequence of observations are independent. 

In this paper, we propose a new nonparametric procedure to detect change points in the mean of high-dimensional time series data. Let $\{X_i \in \mathbb{R}^p, 1\le i \le n\}$ be a sequence of $p$-dimensional observations and $\mu_i$ be the mean of $X_i$ for $i=1, \cdots, n$, where the dimension $p$ can be much larger than the sample size $n$. We first test
\begin{eqnarray}
&H_0&: \mu_1=\cdots=\mu_n, \qquad \mbox{against} \nonumber\\
&H_1&: \mu_1=\cdots=\mu_{\tau_1}\ne \mu_{\tau_1+1}=\cdots=\mu_{\tau_q}\ne \mu_{\tau_q+1}=\cdots=\mu_n, \label{Hypo}
\end{eqnarray}
where $1 \le \tau_1 < \cdots < \tau_q <n$ are some unknown change points. When $H_0$ is rejected, we further estimate the locations of change points. Different from Chen and Zhang (2015) which assumed a sequence of observations to be independent, our procedure incorporates both spatial and temporal dependence, namely spatial dependence among the $p$-components of $X_i$ at each $i$ and temporal dependence between any $X_i$ and $X_j$ for $i \ne j$.   
Different from Bai (2010) which imposed growth rate of the dimension $p$ with respect to the sample size $n$, our procedure allows the dimension $p$ to be much larger than the number of observations $n$. 
Most importantly, our procedure is able to detect a change point on the boundary when data dependence is present. This feature distinguishes our procedure from other existing methods. 
The implementation of the proposed procedure is provided in the R package HDcpDetect (Okamoto et al., 2018).

\setcounter{section}{2} \setcounter{equation}{0}
\section*{\large 2. \bf Main Results}

\subsection*{2.1 \bf Test statistic}
For any $t \in \{1, \cdots, n-1\}$, we consider a bias-corrected statistic
\be
{L}_t=\frac{t(n-t)}{n^2}(\bar{X}_{\le t}-\bar{X}_{>t})^T(\bar{X}_{\le t}-\bar{X}_{>t})-\frac{f_t^T F_{n,M}^{-1} {V}}{n}. \label{estimator}
\ee
Here,  $\bar{X}_{\le t}=t^{-1}\sum_{i=1}^t X_i$ and $\bar{X}_{> t}=(n-t)^{-1}\sum_{i=t+1}^n X_i$. With $M$ defined in Condition 1 of Section 2.2, $f_t$  is an $(M+1)$-dimensional vector with $f_t(1)=1$ and for $i \in \{2, \cdots, M+1\}$,
\begin{eqnarray}
f_t(i)&=&2\biggl\{\frac{(n-t)(t-i+1)}{nt}\mbox{I}(t+1>i)+\frac{t(n-t-i+1)}{n(n-t)}\mbox{I}(n-t+1>i) \nonumber\\
&-&\frac{1}{n}{\sum_{l=1}^{i-1}\mbox{I}(t\ge l)\mbox{I}(n-t\ge i-l)}\biggr\}.\label{vec}
\end{eqnarray}
The element at $i$th row and $j$th column of the $(M+1) \times (M+1)$ matrix $F_{n, M}$ is  
\begin{eqnarray}
F_{n,M}(i, j)&=&(1-\frac{i-1}{n})\mbox{I}(i, j)+(1-\frac{i-1}{n})(1-\frac{j-1}{n})\frac{2-\mbox{I}(j, 1)}{n}\nonumber\\
&-&\frac{1}{n^2}\sum_{a=1}^{n-i+1}\sum_{b=1}^n \biggl\{\mbox{I}(|a-b|+1,j)+\mbox{I}(|a+i-1-b|+1, j)\biggr\}. \label{trans_mat}
\end{eqnarray}
The $i$th component of the $(M+1)$-dimensional random vector $V$ is   
\be
V_i=\frac{1}{n}\sum_{h=1}^{n-i+1}(X_h-\bar{X})^T(X_{h+i-1}-\bar{X}).\label{V.vec}
\ee

Imposing ${n^{-1}f_t^T F_{n,M}^{-1} {V}}$ in (\ref{estimator}) leads to (\ref{mean}) in Proposition 1 that excludes the interference of data dependence in testing the hypothesis and estimating locations of change points in (\ref{Hypo}). The proposed ${L}_t$ depends on $M$ which separates dominant dependence from the remainder. How to choose a proper $M$ in practice will be addressed in Section 3. 
From here to the end of Section 2, we simply assume $M$ to be known in order to present theoretical results of our methods.

For the two-sample testing problem of means,  
${L}_{t}$ can be reduced to the test statistic in Bai and Saranadasa (1996) with temporally independent sequence, and the test statistic in Ayyala et al. (2017) with $m$-dependent Gaussian process. Their asymptotic testing procedures require $t=O(n)$  and thus cannot test the hypothesis in (\ref{Hypo}) if a change point occurs near the boundary, specially at $1$ or $n-1$. Unlike Bai and Saranadasa (1996) and Ayyala et al. (2017), we establish the asymptotic normality of $L_t$ at any $t \in \{1, \cdots, n-1\}$ and the testing procedure can be applied regardless of locations of change points.    
Moreover, there is no need to estimate a change point in the two-sample testing problem as two samples have been pre-specified before testing. In addition to hypothesis testing,  we establish an estimating procedure based on $L_t$ for the locations of change points regardless of locations of change points.

\subsection*{2.2 \bf Hypothesis testing}

To study asymptotic properties of $L_t$, we model the sequence of $p$-dimensional random vectors $\{X_i, 1\le i \le n\}$ by 
\be
X_{i}=\mu_i+\Gamma_i Z \qquad \mbox{for} \quad i=1, \cdots, n,   \label{model}
\ee
where $\mu_i$ is the $p$-dimensional population mean, $\Gamma_i$ is a $p \times q$ matrix with $q\ge n\cdot p$, and $Z=(z_{1}, \cdots, z_{q})^{T}$ so that $\{z_l\}_{l=1}^q$ are mutually independent and satisfy $\mbox{E}(z_l)=0$, $\mbox{var}(z_l)=1$ and $\mbox{E}(z_l^4)=3+\beta$ for some finite constant $\beta$.  

By allowing $\Gamma_i$ to depend on $i$, each $X_i$ has its own covariance described by $\Gamma_i \Gamma_i^T$, and each pair of $X_i$ and $X_j$ has its own temporal dependence described by $\Gamma_i \Gamma_j^T$ for $i \ne j$. Model (\ref{model}) is thus flexible for many applications. We require $q\geq np$ to guarantee the positive definite of $\Gamma_i \Gamma_i^T$. It also ensures the existence of $\Gamma_i$'s under special structural assumptions. For example, if all $X_i$'s are temporally independent, the condition $q\geq np$ guarantees the existence of $\Gamma_i$'s so that $\Gamma_i'\Gamma_j=0$ if $i\neq j\in \{1,\cdots,n\}$.
Another advantage of (\ref{model}) is that it does not assume Gaussian distribution of $Z$ beyond the existence of fourth moment.

Let $C(j-i)=C^T(i-j)=\Gamma_i \Gamma_j^T$, and define a weight function $w_t(h)=\sum_{i=1}^{n-h} n(n-t)\{t^{-1}\mbox{I}(i \le t)-(n-t)^{-1} \mbox{I}(i >t)\}\{t^{-1}\mbox{I}(i+h \le t)-(n-t)^{-1} \mbox{I}(i+h >t)\}$. Moreover, for any matrix $A$, we let $A^{\otimes 2}=AA^T$. 

{\bf Condition1} (Spatial and temporal dependence assumption). We assume that 
$C(i-j)=C(h)$ for $h=i-j$. Moreover, as $n \to \infty$, there exists $M=o(n^{1/2})$ such that
\begin{eqnarray}
&\quad&\sum_{h=M+1}^{n-1} \left|\mbox{tr} \{C(h) \}\right|=o(n), \,\,\mbox{tr}[\{\sum_{h=M+1}^{n-1}w_t(h) C(h)\}^{\otimes 2}]=o(\mbox{tr}[\{\sum_{h=1}^{M} w_t(h) C(h)\}^{\otimes 2}]).\nonumber
\end{eqnarray}

{\bf Condition2} (Covariance assumption). For $h_1$, $h_2$, $h_3$, $h_4 \in \mathcal{A}$ with $\mathcal{A}=\{0, \pm 1, \cdots, \pm M \}$,
\[
\mbox{tr}\{C(h_1) C(h_2) C(h_3) C(h_4)\}=o\biggl[\mbox{tr}\{C(h_1)C(h_2)\}\mbox{tr}\{C(h_3)C(h_4)\} \biggr].
\]

Condition 1 assumes the stationary on $C(i-j)$ 
which can be relaxed to the locally stationary. 
{Condition 1 is trivially true for temporally independent or $m$-dependent sequence}, but general as the sequence needs not be $m$-dependent. 
Moreover, Condition 1 does not impose any structural assumption on dependence within a critical value $M=o(n^{1/2})$, but only requires that the spatial dependence beyond the critical value $M$ is not too strong, so that the two equations are satisfied. At last, comparing to the usually assumed mixing condition, it is advantageous as mixing condition is hard to verify for the real data and usually requires additional smoothness or restrictive moment assumptions (Carrasco and Chen, 2002). 

Condition 2 is imposed on the covariance matrix of the entire sequence of $X_1, \cdots, X_n$. 
To see this, let $X=(X_1^T, X_2^T, \cdots, X_n^T)^T$ and $\Gamma=(\Gamma_1^T, \Gamma_2^T, \cdots, \Gamma_n^T)^T$ from (\ref{model}). The $np \times np$ covariance matrix of $X$ is $\Sigma=\Gamma \Gamma^T$, 
where each $p \times p$ block diagonal matrix of $\Sigma$ describes the spatial dependence among $p$ components of each $X_i$, and each block off-diagonal matrix measures the spatio-temporal dependence of $X_i$ and $X_j$ for $i \ne j$. To impose a condition on $\Sigma$, we may consider $\mbox{tr}(\Sigma^4)=o\{\mbox{tr}^2(\Sigma^2)\}$, which is satisfied if all the eigenvalues of $\Sigma$ are bounded or the dependence of $\Sigma$ is not too strong. However, it is more desirable to impose the condition on the spatial and temporal dependence through $\Gamma_i$. By the relationship that $\Sigma=\Gamma \Gamma^T=(\Gamma_1^T, \Gamma_2^T, \cdots, \Gamma_n^T)^T(\Gamma_1^T, \Gamma_2^T, \cdots, \Gamma_n^T)$, it can be shown that Condition 2 is a sufficient condition for $\mbox{tr}(\Sigma^4)=o\{\mbox{tr}^2(\Sigma^2)\}$.
Another advantage of Condition 2 is that we do not require any explicit relationship between dimension $p$ and the number of observations $n$. 

The mean and variance of ${L}_t$ are given by the following proposition.

\begin{proposition}
\label{proposition1} Under (\ref{model}) and Condition 1, and for $t \in \{1, \cdots, n-1\}$,
\be
\mbox{E}({L}_t)=\frac{t(n-t)}{n^2}(\bar{\mu}_{\le t}-\bar{\mu}_{>t})^T(\bar{\mu}_{\le t}-\bar{\mu}_{>t})-\frac{f_t^T F_{n,M}^{-1}V_B}{n}+o(1), \label{mean}
\ee
where $\bar{\mu}_{\le t}=t^{-1}\sum_{i=1}^t \mu_i$, $\bar{\mu}_{> t}=(n-t)^{-1}\sum_{i=t+1}^n \mu_i$ and $V_B=\{n^{-1}\sum_{i=1}^n(\mu_i-\bar{\mu})^T(\mu_i-\bar{\mu}), \cdots, n^{-1}\sum_{i=1}^{n-M}(\mu_i-\bar{\mu})^T(\mu_{i+M}-\bar{\mu})\}^T$ with $\bar{\mu}=n^{-1}\sum_{i=1}^n\mu_i$.
\begin{eqnarray}
\mbox{var}(L_t)&=&\sigma_{nt}^2=\frac{1}{n^4}\biggl[\sum_{i=1}^n\sum_{j=1}^n \sum_{h_1,h_2\in \mathcal{A}}\{B_t(i, j) B_t(i+h_2, j-h_1)+B_t(i, j) B_t(j-h_1, i+h_2)\}\nonumber\\
&\times& \mbox{tr}\{C(h_1)C(h_2)\}+\sum_{i=1}^n\sum_{j=1}^n \sum_{k=1}^n \sum_{h\in \mathcal{A}\cup \mathcal{A}^c}\{B_t(i,j)
+B_t(j,i)\}\{B_t(k, i+h)\nonumber\\
&+&B_t(i+h,k) \}\mu_j^T C(h)\mu_k\biggr]\{1+o(1)\}, \label{variance}
\end{eqnarray}
where the set $\mathcal{A}=\{0, \pm 1, \cdots, \pm M \}$ and the $n\times n$ matrix $B_t$ satisfies
\begin{align*}
B_t(i, j)&=\frac{n-t}{t} \mbox{I}(i \le t) \mbox{I}(j \le t)-{2} \mbox{I}(i \le t) \mbox{I}(j > t)+\frac{t}{n-t} \mbox{I}(i > t) \mbox{I}(j > t)\\
&-\sum_{h=0}^{M} (f_t^T F^{-1}_{n,M})_{h+1}\big\{{\mbox{I}(i-j, h)} -\frac{\mbox{I}(j \ge h+1)+\mbox{I}(j \le n-h)}{n}+\frac{n-h}{n^2}\big\}.
\end{align*}
\end{proposition}

Now we are ready to present the asymptotic normality of ${L}_t$. 

\begin{theorem}
\label{theorem1} Assume (\ref{model}) and Conditions 1--2. As $n \to \infty$ and for any $t \in \{1, \cdots, n-1\}$,
$\{{L}_t-\mbox{E}({L_t})\}/{\sigma_{nt}}$ converges in distribution to the standard normal $N(0,1)$,  
where $\sigma_{nt}$ is defined by (\ref{variance}) in Proposition 1.
\end{theorem}

To implement a testing procedure based on Theorem 1, we need to estimate 
\[
{\sigma}_{nt,0}^2=\sum_{i, j=1}^n \sum_{h_1,h_2\in \mathcal{A}}\frac{B_t(i, j)}{n^4}\{ B_t(i+h_2, j-h_1)+B_t(j-h_1, i+h_2) \} {\mbox{tr}\{C(h_1)C(h_2)\}},
\]
which is $\mbox{var}({L}_t)$ under the null hypothesis.
The only unknown terms are $\mbox{tr}\{C(h_1)C(h_2)\}$ for $h_1$ and $h_2$ from $\mathcal{A}=\{0, \pm 1, \cdots, \pm M \}$. Similar to Li and Chen (2012), we estimate them by
\begin{eqnarray}
T_{est}&=&\frac{1}{n_1^*}\sum_{s,t}^*X_{t+h_2}^T X_s X_{s+h_1}^T X_t-\frac{1}{n_2^*}\sum_{r,s,t}^* X_r^T X_s X_{s+h_1}^T X_t
-\frac{1}{n_3^*}\sum_{r,s,t}^* X_r^T X_s X_{s+h_2}^T X_t\nonumber\\
&+&\frac{1}{n_4^*}\sum_{q,r,s,t}^* X_q^T X_r X_{s}^T X_t, \label{var.est}
\end{eqnarray}
where $\sum^*$ represents the sum of indices that are at least
 $M$ apart, and $n_i^*$ with $i=1, 2, 3, 4$ are the corresponding number of indices. As a result, the estimator of ${\sigma}_{nt,0}^2$ is
\begin{eqnarray}
{s}_{t}^2&=&\sum_{i, j=1}^n\sum_{h_1,h_2\in \mathcal{A}}\frac{B_t(i, j)}{n^4} \biggl\{ B_t(i+h_2, j-h_1)
+B_t(j-h_1, i+h_2) \biggr\} T_{est}.\label{var.est2}
\end{eqnarray}

\begin{figure}[t!]
\begin{center}
\includegraphics[width=0.32\textwidth,height=0.32\textwidth]{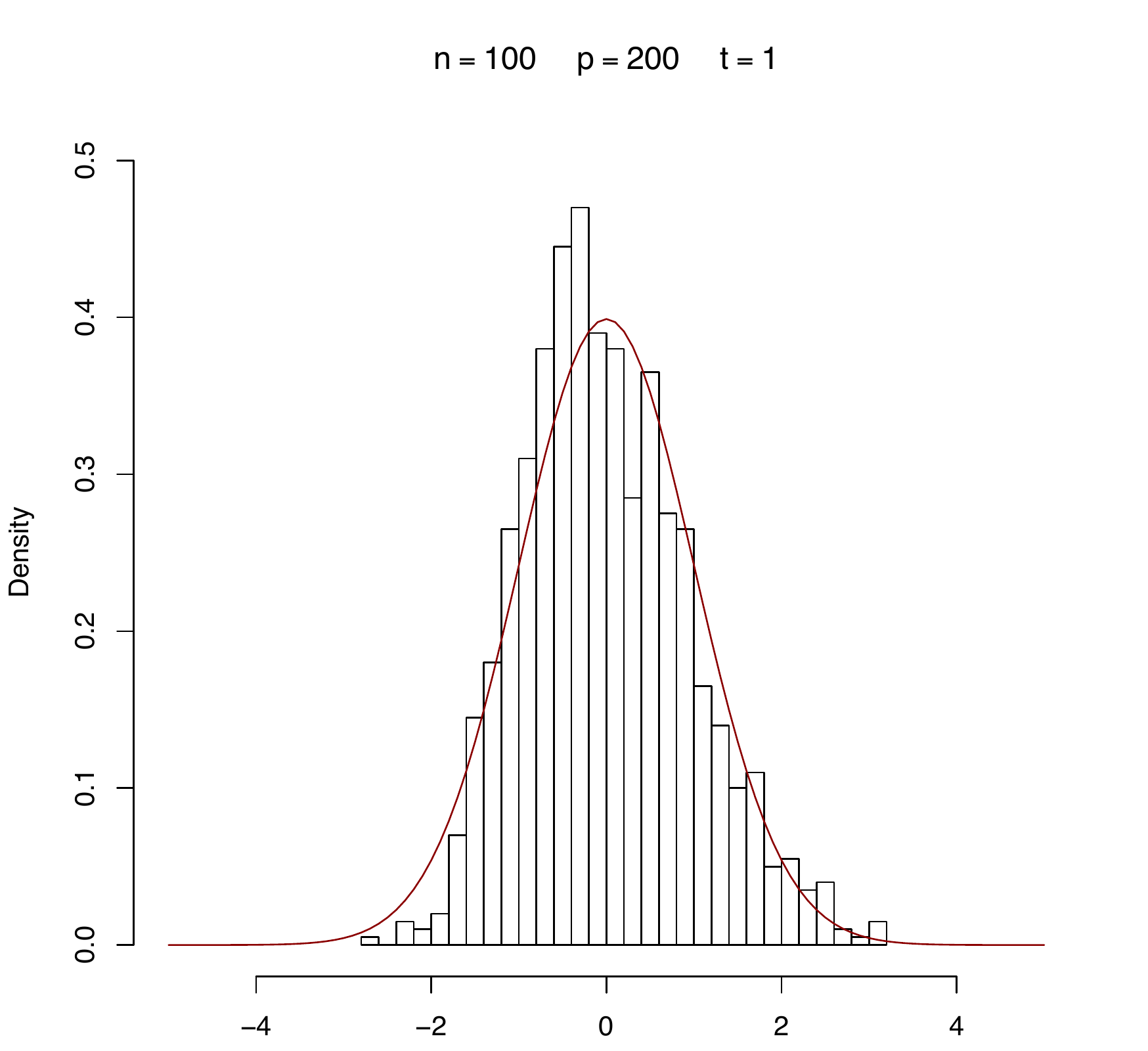}
\includegraphics[width=0.32\textwidth,height=0.32\textwidth]{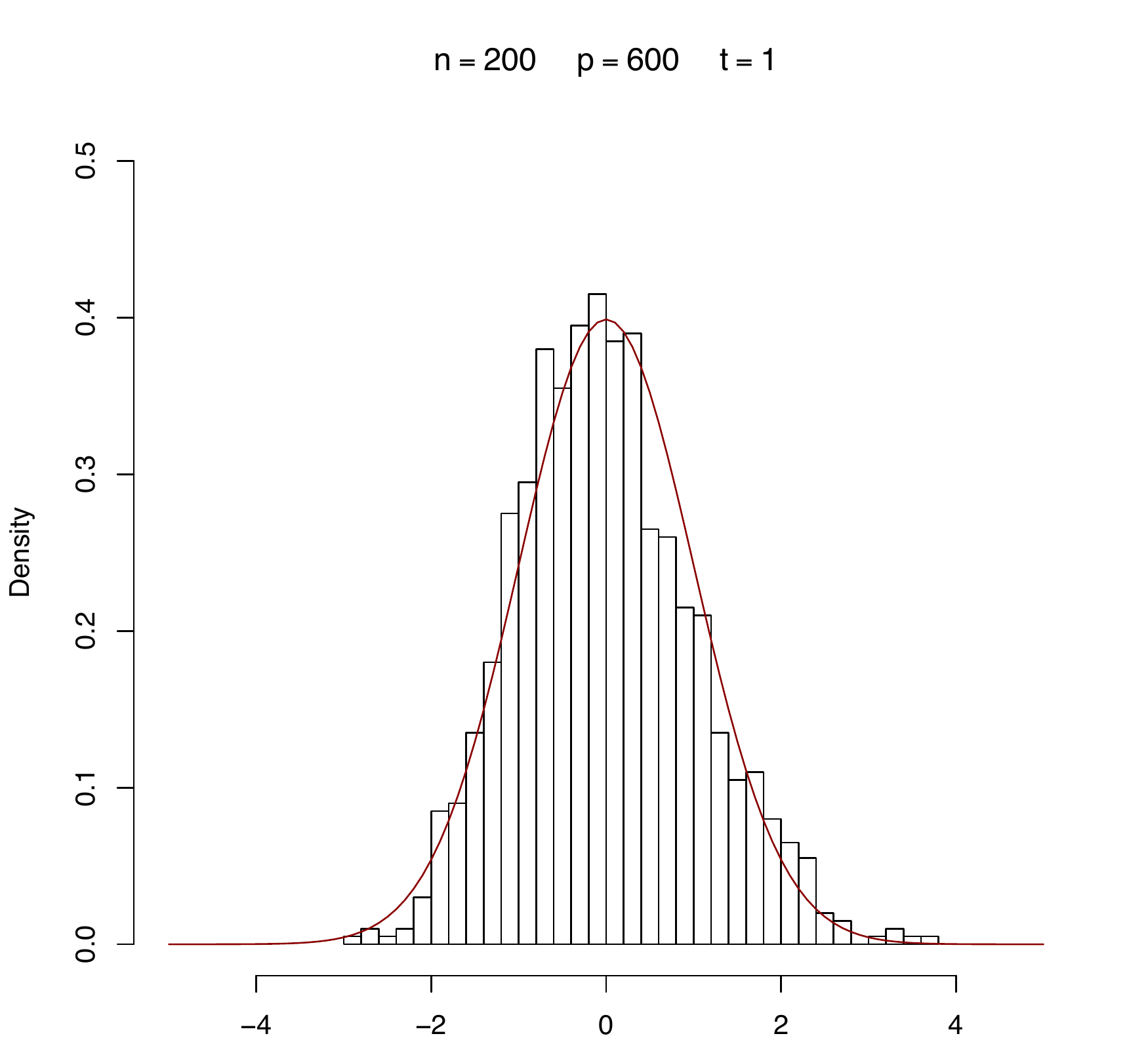}
\includegraphics[width=0.32\textwidth,height=0.32\textwidth]{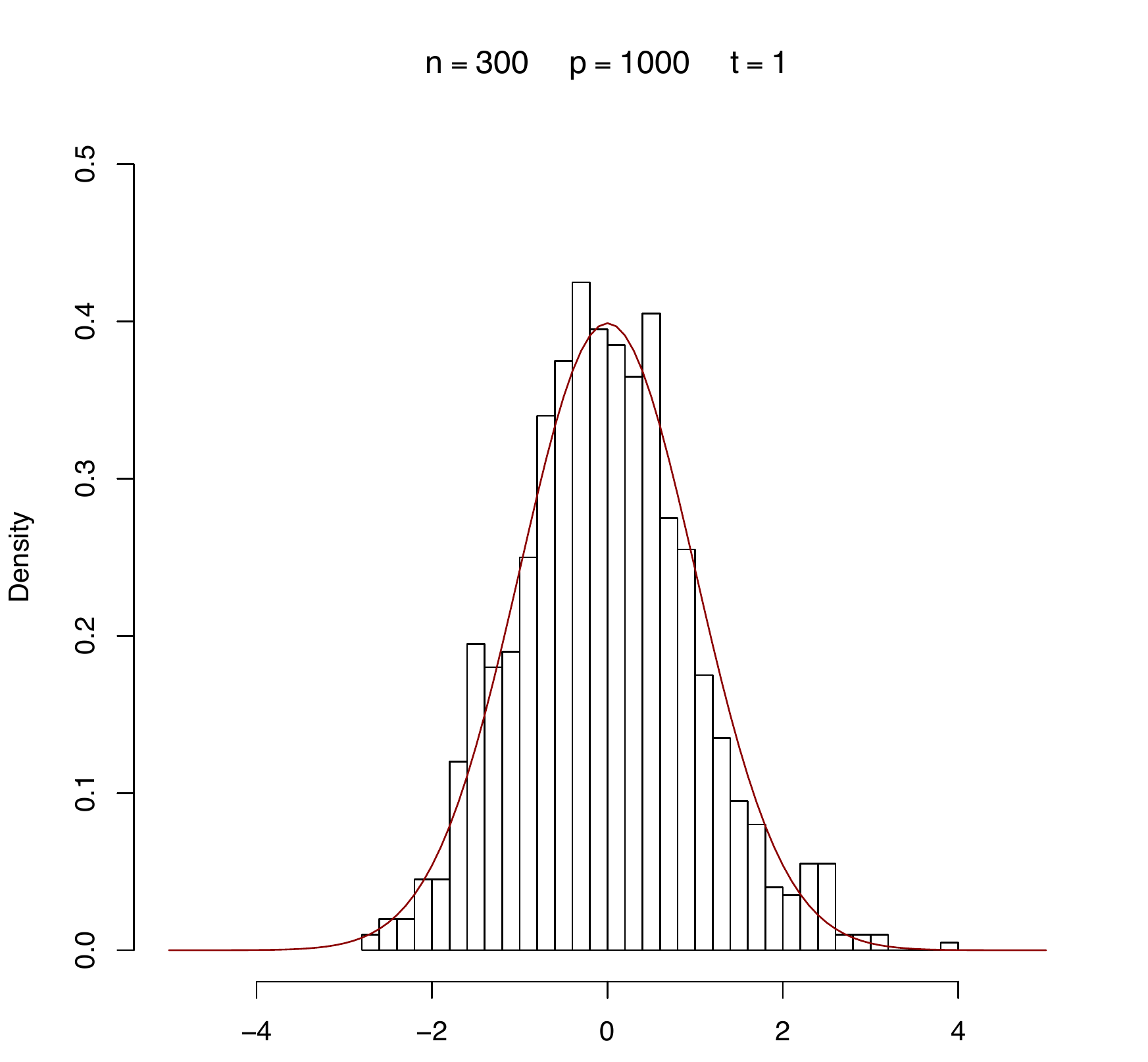}\\
\includegraphics[width=0.32\textwidth,height=0.32\textwidth]{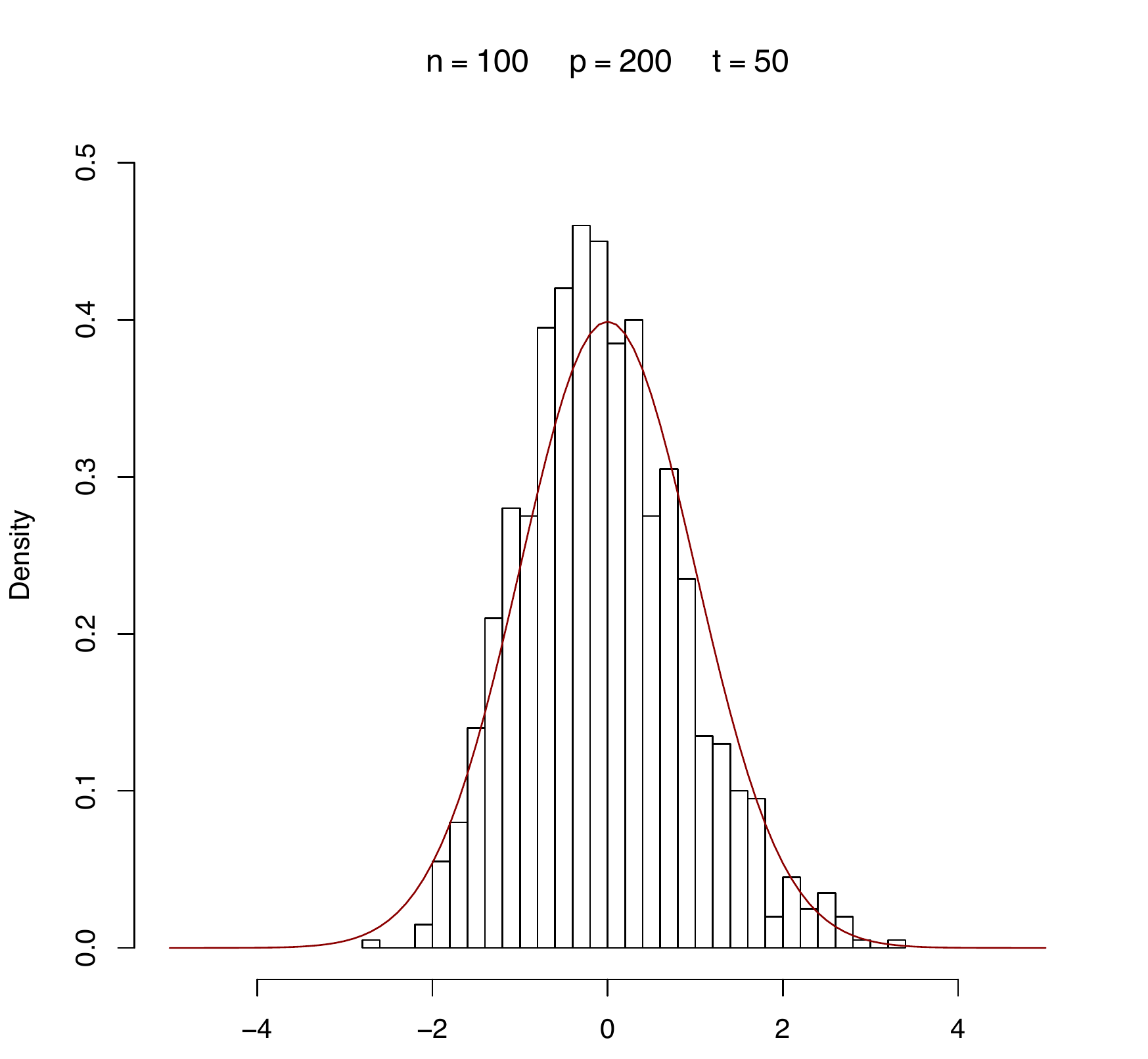}
\includegraphics[width=0.32\textwidth,height=0.32\textwidth]{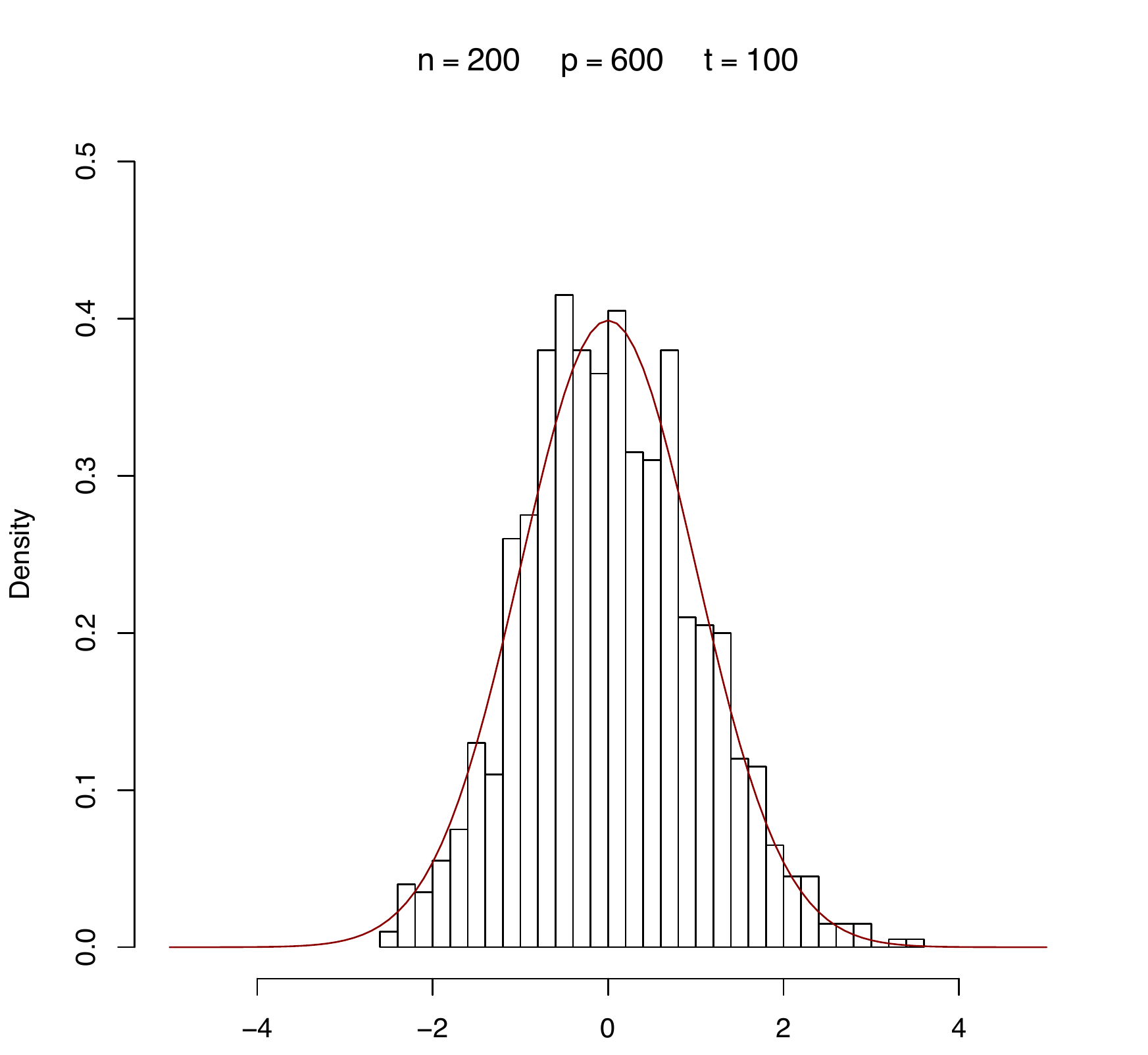}
\includegraphics[width=0.32\textwidth,height=0.32\textwidth]{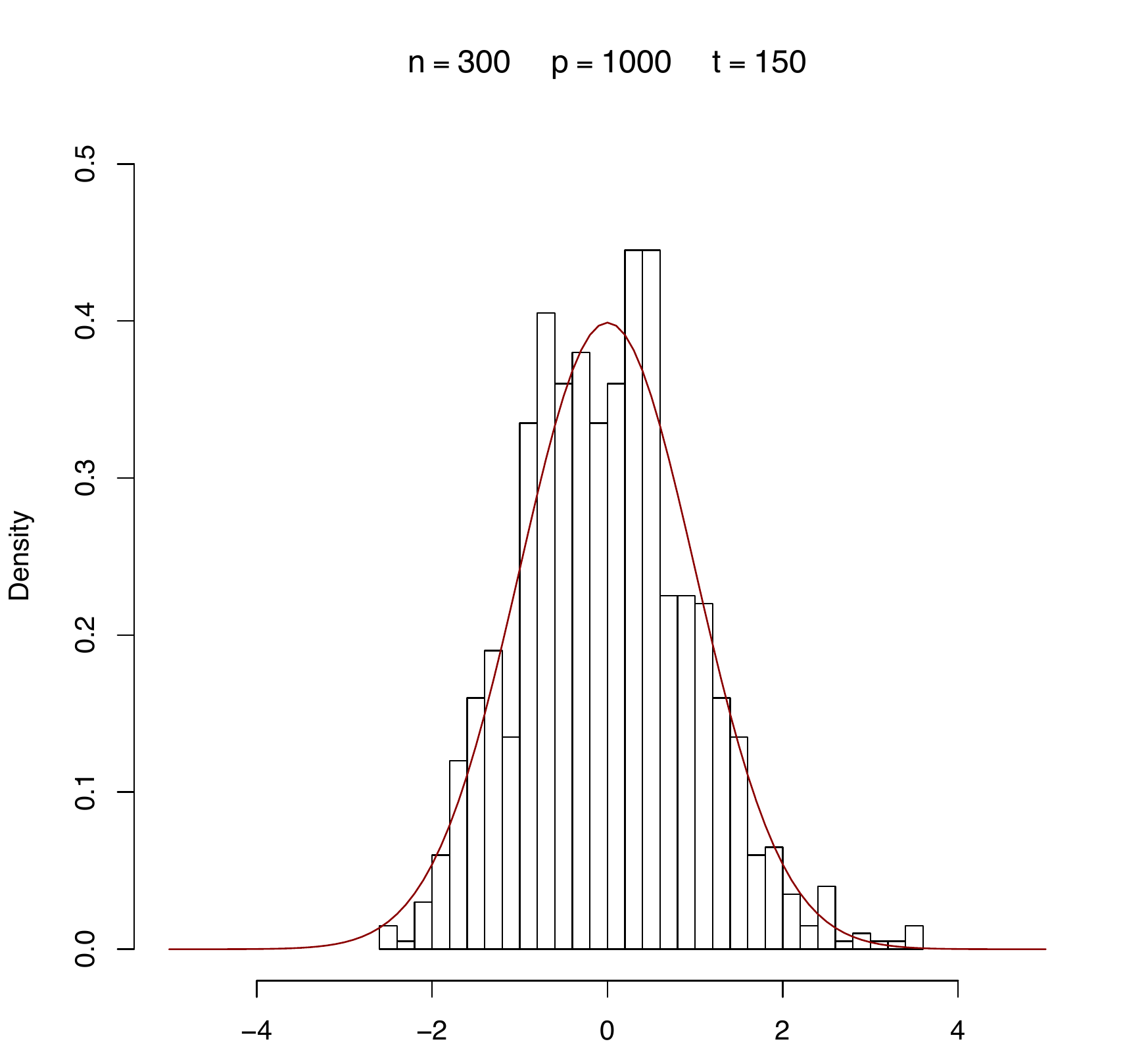}
\caption{Histogram of ${L}_t/{s}_{t}$ versus $\mbox{N}(0,1)$-curve. The upper row chooses $t=1$, at different $n$ and $p$; The lower row chooses $t=n/2$, and different $n$ and $p$. }
\label{fig1}
\end{center}
\end{figure}

\begin{theorem}
\label{theorem2} Assume the same conditions in Theorem 1 and $H_0$ of (\ref{Hypo}). As $n \to \infty$ and for any $t \in \{1, \cdots, n-1\}$,
${L}_t/{s_{t}}$ converges in distribution to the standard normal $N(0,1)$. 
\end{theorem}


One of the contributions in this paper is establishing the asymptotic normality of $L_t$ for any $t \in \{1, \cdots, n-1\}$. This enables us to test the hypothesis of (\ref{Hypo}) even when a change point is on the boundary of a sequence.    
We conduct some simulations for a visual inspection. Figure \ref{fig1} illustrates histograms of ${L}_t/{s}_{t}$ based on $1000$ iterations for $t=1$ and $t=n/2$, respectively. The data were generated based on the setups in Section 4.1. Clearly, as $n$ and $p$ increase, the histograms converge to the standard normal curve even when $t$ equals 1.

From Theorem 2, our testing procedure rejects $H_0$ of (\ref{Hypo}) if ${L}_t/{s}_{t}> z_{\alpha}$ with a nominal significance level $\alpha$ , where $z_{\alpha}$ is the upper-$\alpha$ quantile of $N(0,1)$. The testing procedure relies on $t$ and may lose power if the chosen $t$ is very different from the location of a change point. For example, there exists only one change point located near the boundary. If we choose a $t$ near the middle to break the entire sequence into two subsequences, the small piece with mean change falls into a long subsequence and its contribution to the change point detection is diluted by averaging all observations in the subsequence. 
In order to circumvent the difficulty of choosing $t$ and most importantly retain the power of the test, we accumulate all the marginal ${{L}_t}$ and consider 
\begin{eqnarray}
{\mathcal{L}}= \sum_{t=1}^{n-1} {{L}_t}. \label{L2stat}
\end{eqnarray}

Let $\mathcal{B}(i, j)= \sum_{t=1}^{n-1} B_t(i, j)$ where $B_t(i, j)$ is specified in Proposition 1, $\sigma_n^2$ be $\mbox{var}({\mathcal{L}})$ obtained by replacing $B_t(\cdot, \cdot)$ with $\mathcal{B}(\cdot, \cdot)$ in (\ref{variance}), and $s^2$ be the estimator of $\mbox{var}({\mathcal{L}})$ under the null hypothesis obtained by replacing $B_t(\cdot, \cdot)$ with $\mathcal{B}(\cdot, \cdot)$ in (\ref{var.est2}).

\begin{theorem}
\label{theorem3} Assume the same conditions in Theorem 1. As $n \to \infty$,
$\{{\mathcal{L}}-\sum_{t=1}^{n-1}\mbox{E}(L_t)\}/{\sigma_{n}}$ converges  to the standard normal in distribution. 
Especially under $H_0$ of (\ref{Hypo}),
${{\mathcal{L}}}/{s}$ converges to the standard normal in distribution.
\end{theorem}

Based on Theorem 3, 
we reject $H_0$ of (\ref{Hypo}) with a nominal significance level $\alpha$ if ${{\mathcal{L}}}/{s}> z_{\alpha}$. Free of the tuning parameter $t$ and retaining the power, the testing procedure based on ${\mathcal{L}}$ is thus chosen for the existence of any change point.


{\bf Remark 1.} Alternatively, one may consider the max-norm statistic $\max_{1 \le t \le n-1} {{L}_t}/{s_t}$. 
If there are only few time points $t$ where the difference of $\bar{\mu}_{\le t}$ and $\bar{\mu}_{>t}$ is large, the max-norm based test is expected to be more powerful than our proposed test. If the small differences occur in many time points, our test can dominate the max-norm based test  {by aggregating all  small differences}. 
Furthermore, it requires stringent conditions to establish the extreme value distribution of $\max_{1 \le t \le n-1} {{L}_t}/{s_t}$ and its rate convergence is known to be slow (Liu and Shao, 2013).


\subsection*{2.2 \bf Estimating one and multiple change points}

If the null hypothesis is rejected, we further estimate the change points. We first consider the case of one change point. The location of a change point $\tau \in \{1, \cdots, n-1\}$ is estimated by
\be
{\tau}_{e}= \argmax_{0< t/n < 1} {L}_t, \label{identification}
\ee
where ${L}_t$ is given by (\ref{estimator}). 
The rationale of proposing ${\tau}_{e}$ is demonstrated by Lemma 1. 

\begin{lemma}
\label{lemma1} Under (\ref{model}), Condition 1 and $H_1$ of (\ref{Hypo}), $\mbox{E}(L_t)$ always attains its maximum at the change point $\tau \in \{1, \cdots, n-1\}$. 
\end{lemma}

Let $\delta^2=(\mu_1-\mu_n)^T(\mu_1-\mu_n)$ and $v_{\max}=\sqrt{ \max_{0<t<n} n^2 \sigma_{nt}^2}$ where $\sigma_{nt}^2$ is given in Proposition 1. Here $\delta^2$ and $v_{\max}$ measure signal strength and maximal noise, respectively. The following theorem establishes the convergence rate of ${\tau}_e$.

\begin{theorem}
\label{theorem4} Assume that the change-point $\tau \in \{1, \cdots, n-1\}$ satisfies $\min \{\tau, n-\tau\}=O(n^{\gamma})$ with $\gamma \in [0, 1]$. 
Under the same conditions in Theorem 1, as $n \to \infty$,
\[
{\tau}_e-\tau=O_p\biggl (\frac{{n^{1-\gamma}{\mbox{log}^{1/2}n}}\,\,{v}_{\max}}{\delta^2} \biggr).
\]
\end{theorem}

{\bf Remark 2.}  
Under cross-sectional dependence but temporal independence, 
we can derive
$$v_{max}=\sqrt{2\mbox{tr}\{C^2(0)\}+4n^{-1}\max_{0<t<n}t(n-t) (\bar{\mu}_{\le t}-\bar{\mu}_{>t})^TC(0)(\bar{\mu}_{\le t}-\bar{\mu}_{>t})}.$$    
Moreover, under the local alternative that the change in $\mu$ tends to zero, the leading order $v_{max}=\sqrt{2\mbox{tr}\{C^2(0)\}}=O(p^{1/2})$ if all the eigenvalues of $C(0)$ are bounded, and thus 
\[
{\tau}_e-\tau=O_p\biggl(\frac{n^{1-\gamma}\mbox{log}^{1/2}n\,\,p^{1/2}}{\delta^2} \biggr).
\] 

{\bf Remark 3.} In the change point literature, it commonly assumes that the change point $\tau$ is of the form $\kappa n$ with $\kappa \in (0, 1)$, {that is $\tau=O(n)$ with $\gamma=1$ in terms of our notation}. 
The corresponding convergence rate is ${\mbox{log}^{1/2}n\,\,p^{1/2}}{\delta^{-2}}$. This excludes the case that the change point is near or on the boundary. 
{Theorem 4 is general as $\gamma$ can vary within [0, 1]. Especially, when the change point $\tau=O(1)$} (near or on the boundary), the convergence rate is  $n{\mbox{log}^{1/2}n\,\,p^{1/2}}{\delta^{-2}}$ which is $n$ times slower than the convergence rate when $\tau=O(n)$.  



To estimate the locations of multiple change points $1 \le \tau_1 < \cdots < \tau_q <n$, we can iteratively apply a binary segmentation method similar to that in Vostrikova (1981). Suppose that we have already estimated $l-1$ change points as $1 \le {\tau}_{e, 1}< \cdots < {\tau}_{e,  l-1} <n$, which partition the original data into $l$ segments. Define ${\tau}_{e,0}=0$ and ${\tau}_{e,l}=n$. Let ${L}_t[{\tau}_{e, i-1}+1, {\tau}_{e,i}]$, ${\mathcal{L}}[{\tau}_{e, i-1}+1, {\tau}_{e,i}]$ and $s[{\tau}_{e, i-1}+1, {\tau}_{e, i}]$ be the statistics calculated based on data from the $i$th segment $[{\tau}_{e, i-1}+1, {\tau}_{e, i}]$. For each of $l$ segments, we first conduct hypothesis testing by checking if ${\mathcal{L}}[{\tau}_{e, i-1}+1, {\tau}_{e,i}]/s[{\tau}_{e, i-1}+1, {\tau}_{e,i}] \le z_{\alpha_n}$ where $\alpha_n$ is a chosen nominal significance level. If yes, no change point is estimated from $[{\tau}_{e, i-1}+1, {\tau}_{e, i}]$. Otherwise, one change point is estimated as ${\tau}_{e, l^*}=\mbox{arg} \max_{t \in [{\tau}_{e, i-1}+1, {\tau}_{e, i}]} {L}_t[{\tau}_{e, i-1}+1, {\tau}_{e, i}]$, which further partitions $[{\tau}_{e, i-1}+1, {\tau}_{e, i}]$ into $[{\tau}_{e, i-1}+1, {\tau}_{e, l^*}]$ and $[{\tau}_{e, l^*}+1, {\tau}_{e, i}]$. Repeat the above procedure iteratively until no more change point can be estimated from any segment.

Let $\mathbb{S}$ be the set of all change points $\{\tau_1, \cdots, \tau_q \}$ and ${\mathbb{S}_e}$ be the set of estimated change points, respectively. Letting $\tau_0=0$ and $\tau_{q+1}=n$, we define
$$\mbox{SNR}_{\min}=\min_{a+1<b} \frac{\mbox{E}(\mathcal{L}[\tau_a+1,\tau_b ])}{\sigma_n[\tau_a+1,\tau_b ]}$$
to be the minimal signal-to-noise ratio from all segments, each of which has starting point $\tau_a+1$ for $a \in \{0, \cdots, q\}$ and ending point $\tau_b$ for $b \in \{1, \cdots, q+1\}$.
We establish the consistency of ${\mathbb{S}_e}$ under the following Condition 3 plus Conditions 1--2.


{\bf Condition 3} (Minimal signal-to-noise ratio assumption). As $n \to \infty$, $\alpha_n \to 0$ and $\mbox{SNR}_{\min}$ diverges such that $z_{\alpha_n}=o(\mbox{SNR}_{\min})$. Furthermore, in Theorem 4, $v_{\max}[\tau_a+1,\tau_b ]=o\{\delta^2[\tau_a+1,\tau_b ]/(n^{1-\gamma}\mbox{log}^{1/2}n)\}$ for all $[\tau_a+1,\tau_b ]$ that contains at least one change point.

\begin{theorem}
\label{theorem5} Assume (\ref{model}) and Conditions 1--3. As $n \to \infty$,
 ${\mathbb{S}_e}$ converges to $\mathbb{S}$ in probability. 
\end{theorem}

{\bf Remark 4.} The binary segmentation can control the family-wise error rate (FWER) as we set $\alpha_n\to 0$ in Condition 3. Especially, one can choose $\alpha_n=1/\{n\log(n)\}$ so that the FWER is controlled even if other conditions in Theorem 5 are not satisfied. 

{\bf Remark 5.} 
The defined $\mbox{SNR}_{\min}$ provides a quantitative measure for efficiency of the binary segmentation. To appreciate this, we consider a configuration of two change points $\tau_1$ and $\tau_2$. Let $\tau_0=0$, $\tau_{3}=n$. The piecewise constant signals are zero in $[\tau_0, \tau_1]$ and $[\tau_2, \tau_3]$ and positive in $[\tau_1, \tau_2]$. Then $\mbox{SNR}_{\min}$ is  the smallest signal-to-noise ratio from $[\tau_0, \tau_2]$, $[\tau_0, \tau_3]$, $[\tau_1, \tau_3]$. Especially, if $[\tau_1, \tau_2]$ is short and buried in the middle of the large segment $[\tau_0, \tau_3]$ (Olshen and Venkatraman, 2004), $\mbox{SNR}_{\min}$ is close to zero. The binary segmentation is well known to be inefficient under this configuration.
To improve its performance, we may consider ${L}_{t_1, t_2}=\sum_{t=t_1}^{t_2-1} {L}_{t_1, t_2}^t$, where ${L}_{t_1, t_2}^t$ is the test statistic (\ref{estimator}) defined in a randomly generated interval $[t_1, t_2]$ with $1 \le t_1 < t_2 \le n$. The rationale is that when $[t_1, t_2]$ happens to be $[\tau_0, \tau_2]$ or $[\tau_1, \tau_3]$, the change-point detection will be more powerful than that based on the entire sequence. Based on ${L}_{t_1, t_2}$, the circular binary segmentation or wild binary segmentation can be implemented accordingly. 

\setcounter{section}{3} \setcounter{equation}{0}
\section*{\large 3. \bf Elbow Method for Dependence}

The proposed procedure relies on the choice of $M$, which is unknown in practice. From Condition 1, $M$ separates dominant temporal dependence from the remainder.  As demonstrated in simulation studies of Section 4, if data are dependent ($M \ne 0$), wrongly applying the procedure based on the assumption that $M=0$ can cause severe type I error and thus produce a lot of false positives when estimating locations of change points. On the other hand, choosing a value that is larger than the actual $M$  will reduce the power of the test and thus generate more false negatives. 
Here we propose a quite simple way to determine $M$.

\begin{figure}[t!]
\begin{center}
\includegraphics[width=0.45\textwidth,height=0.4\textwidth]{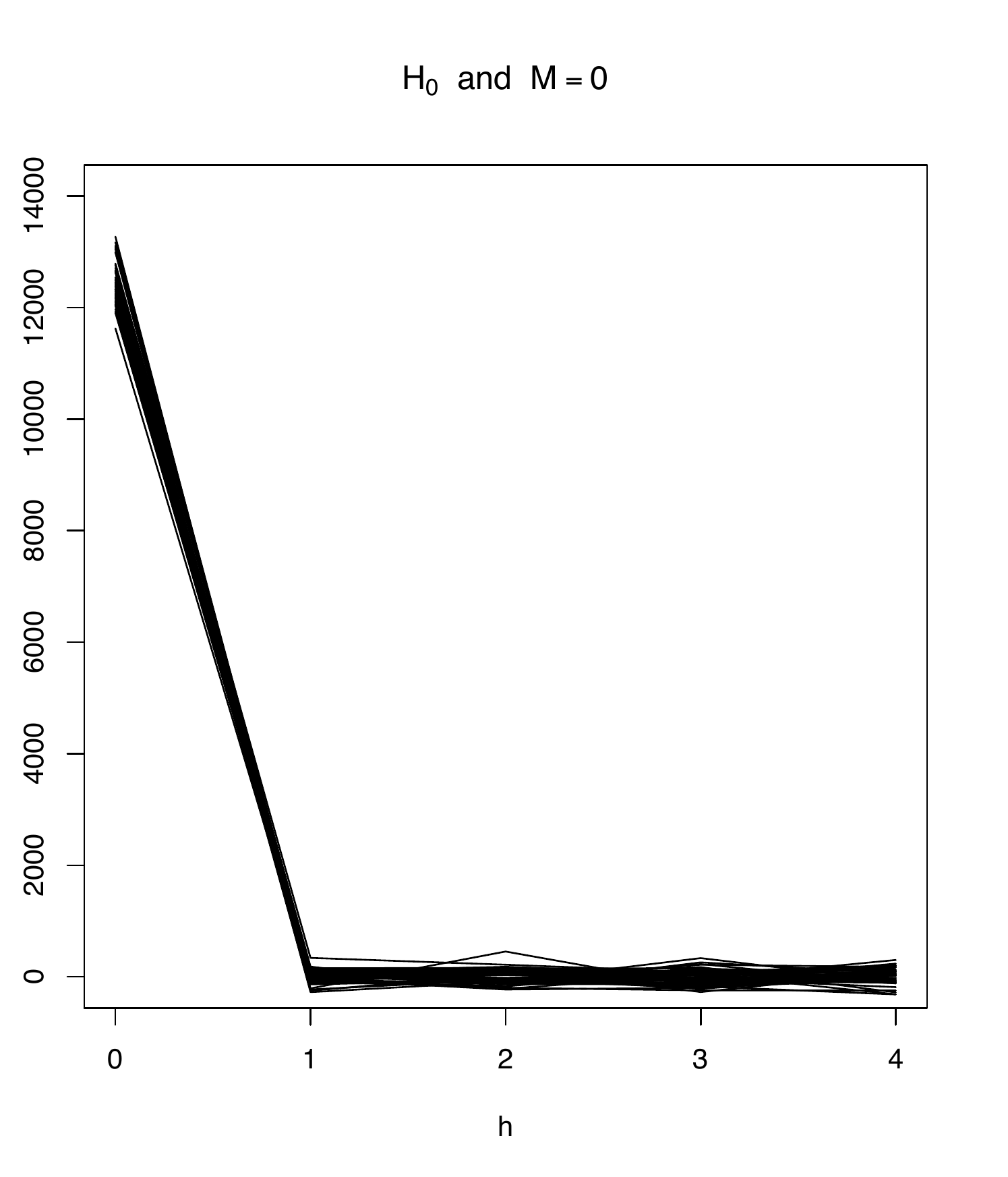}
\includegraphics[width=0.45\textwidth,height=0.4\textwidth]{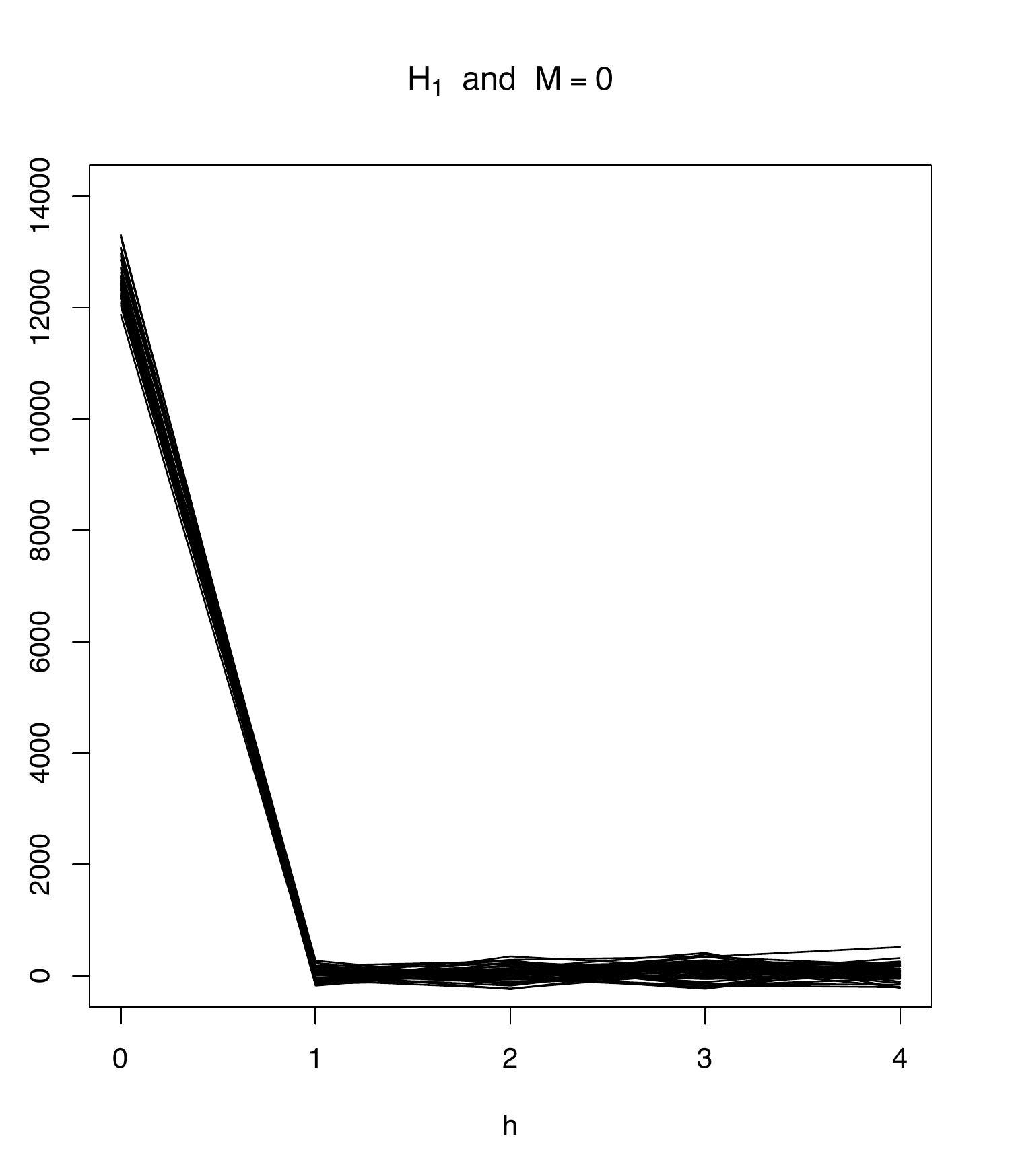}\\
\includegraphics[width=0.45\textwidth,height=0.4\textwidth]{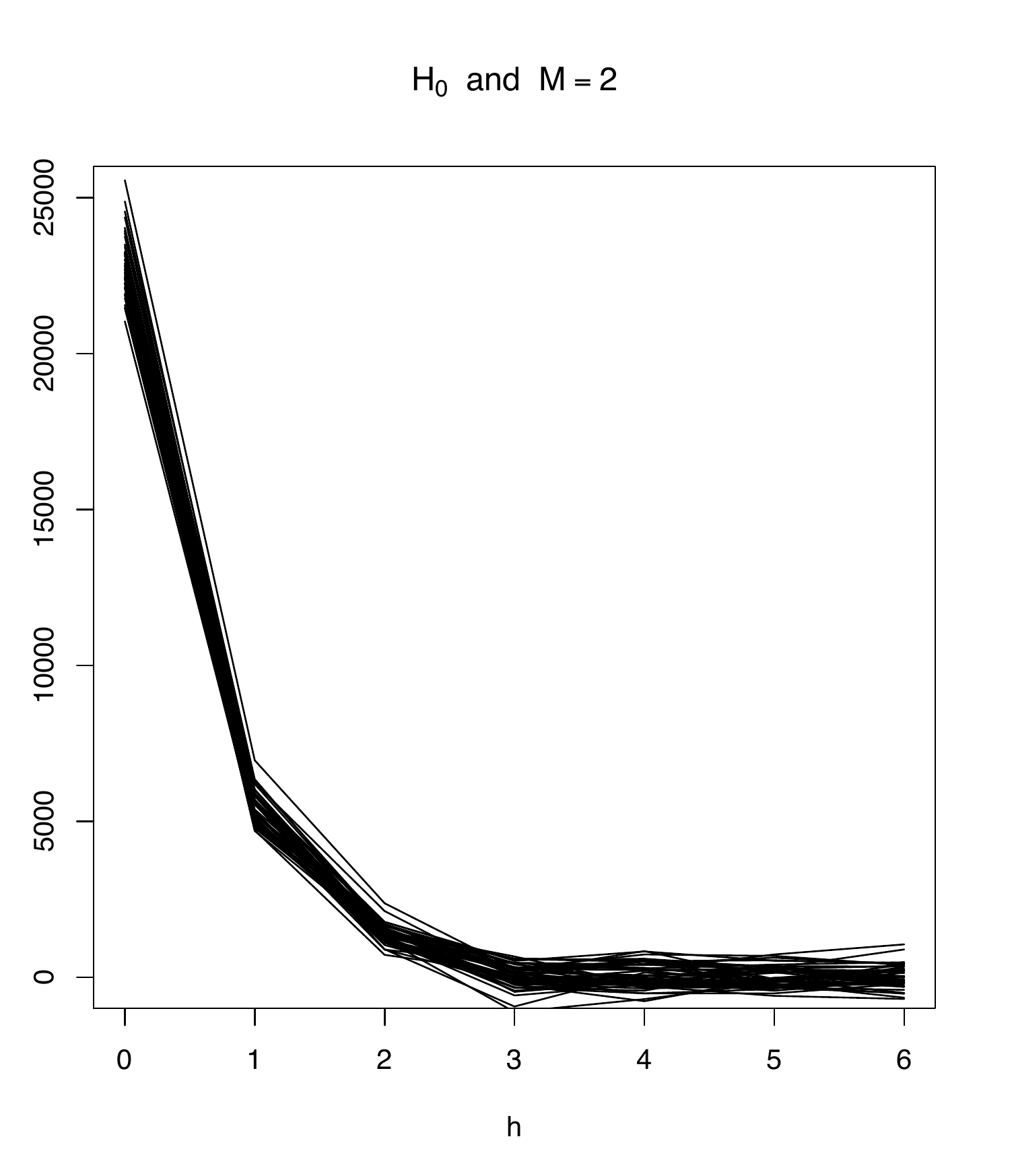}
\includegraphics[width=0.45\textwidth,height=0.4\textwidth]{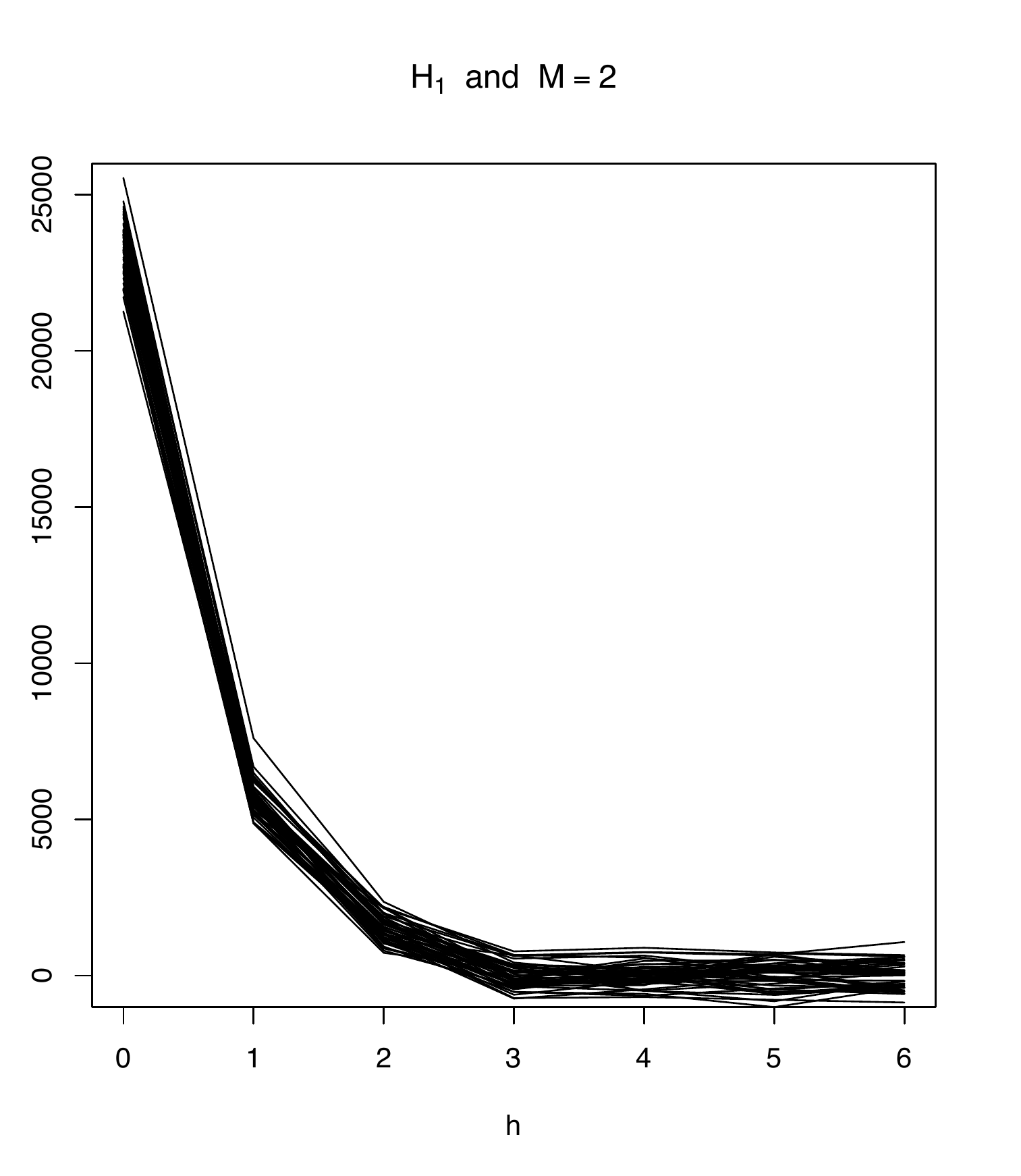}
\caption{The elbow method for choosing $M$ under both null and alternative hypotheses. The results were obtained based on $50$ replications.}
\label{fig5}
\end{center}
\end{figure}

Condition 1 states that $\mbox{Cov}(X_i, X_j)=C(i-j)$ is relatively small if $|i-j| > M$, or equivalently, $\mbox{tr}\{C(h)C^T(h)\}$ is small if $|h| > M$. The unknown $\mbox{tr}\{C(h)C^T(h)\}$ can be consistently estimated by (\ref{var.est}) with $h_1=-h_2=h$ under the null hypothesis according to the proof of Theorem 2. Even under the alternative hypothesis, the effect of heterogeneity of means $\mu_i$ on the estimation is of small order as long as the heterogeneity is not too strong. 
We thus determine $M$ by calculating (\ref{var.est}) for each integer starting from $0$, and terminate the process once a small value appears. Visually, we can plot (\ref{var.est}) versus $h$, and 
use the elbow in the plot to determine $M$. 

To demonstrate the idea above, we generated the random sample $\{X_i \}$ for $i=1, \cdots, n$ using (\ref{data_g}) in Section 4 with $n=150$ and $p=600$. We considered $M=0$ and $2$, respectively. Figure \ref{fig5} illustrates (\ref{var.est}) versus $h$ based on $50$ iterations. When the actual $M=0$, the elbow happened at $h=1$ under both null and alternative hypotheses. We thus estimated $M$ by $0$. Similarly, when $M=2$, the elbow happened at $h=3$ which suggested us to estimate $M$ by $2$.

\setcounter{section}{4} \setcounter{equation}{0}
\section*{\large 4. \bf Simulation Studies}

\subsection*{4.1 \bf Empirical performance of the testing procedure}

\begin{table}[t!]
\tabcolsep 4.8pt
\begin{center}
\caption{Empirical sizes and powers of the CQ, the E-div and the proposed tests based on $1000$ replications with Gaussian $\epsilon_i$ in (\ref{data_g}).}
\label{case1}
\begin{tabular}{ccccccccccccc}
 \multicolumn{13}{c}{Size}\\
 &&\multicolumn{3}{c}{$n=100$} &&\multicolumn{3}{c}{$150$} &&\multicolumn{3}{c}{$200$}\\[1mm]
$M$ &method & $p=200$ & $600$ &$1000$ &  & $200$ & $600$ & $1000$ & & $200$& $600$ & $1000$ \\
     &CQ  & $0.066$ & $0.059$ & $0.066$& & $0.055$ & $0.071$&$0.054$ & & $0.051$& $0.062$ & $0.040$\\
 $0$    &E-div  & $0.069$ & $0.050$& $0.055$& & $0.042$ & $0.047$& $0.045$& & $0.038$& $0.071$&$0.041$ \\
    & New  & $0.056$ & $0.055$ & $0.051$& & $0.052$ & $0.067$&$0.060$ & & $0.052$& $0.062$ &$0.040$ \\
     &CQ  & $0.859$ & $0.999$ & $1.000$& & $0.875$ & $0.999$&$1.000$ & & $0.881$& $0.999$ & $1.000$\\
 $1$    &E-div  & $0.989$ & $1.000$& $1.000$& & $1.000$ & $1.000$& $1.000$& & $1.000$& $1.000$&$1.000$ \\
    & New  & $0.039$ & $0.044$ & $0.053$& & $0.059$ & $0.046$&$0.050$ & & $0.050$& $0.044$ &$0.054$ \\
     &CQ  & $0.990$ & $1.000$ & $1.000$& & $0.991$ & $1.000$&$1.000$ & & $0.995$& $1.000$ & $1.000$\\
 $2$    &E-div  & $1.000$ & $1.000$& $1.000$& & $1.000$ & $1.000$& $1.000$& & $1.000$& $1.000$&$1.000$ \\
    & New  & $0.054$ & $0.034$ & $0.043$& & $0.048$ & $0.054$&$0.039$ & & $0.047$& $0.054$ &$0.043$ \\
\multicolumn{13}{c}{Power}\\
     &CQ  & $0.274$ & $0.318$ & $0.401$& & $0.407$ & $0.524$&$0.601$ & & $0.558$& $0.764$ & $0.826$\\
 $0$    &E-div  & $0.107$ & $0.147$& $0.164$& & $0.142$ & $0.195$& $0.224$& & $0.146$& $0.253$&$0.326$ \\
    & New  & $0.190$ & $0.193$ & $0.234$& & $0.273$ & $0.304$&$0.366$ & & $0.327$& $0.508$ &$0.555$ \\
\end{tabular}
\end{center}
\end{table}

The first part of simulation studies is to investigate the empirical performance of the test statistic $\mathcal{{L}}$ with asymptotic normality established in Theorem 3. 
The random sample $\{X_{i}\}$ for $i=1, \cdots, n$, were generated from the following multivariate linear process
\be
X_{i}=\mu_i+\sum_{l=0}^{M+2} Q_{l}\,\epsilon_{i-l}, \label{data_g}
\ee
where $\mu_i$ is the $p$-dimensional population mean vector at point $i$, $Q_{l}$ is a $p \times p$ matrix for $l=0, \cdots, M+2$, and $\epsilon_{i}$ is a $p$-variate random vector with mean $0$ and identity covariance $I_p$.
In the simulation, we set $Q_{l}=\{0.6^{|i-j|}(M-l+1)^{-1}\}$ for $i, j=1, \cdots, p$, and $l=0,\cdots, M$. For $Q_{M+1}$ and $Q_{M+2}$, we considered two different scenarios. If $M=0$, we simply chose $Q_{M+1}=Q_{M+2}=0$ so that $\{X_i\}_{i=1}^n$ became an independent sequence. If $M \ne 0$, we chose $Q_{M+1}=Q_{M+2}$ and each row of them had only $0.05 p$ non-zero elements that were randomly chosen from $\{1, \cdots, p\}$ with magnitude generated by Unif $(0, 0.05)$. By doing so, the dependence was dominated by $Q_l$ for $l=0, \cdots, M$ plus perturbations contributed by $Q_{M+1}$ and $Q_{M+2}$.

Without loss of generality, we chose $\mu_i=0$ for $i=1, \cdots, n$ under $H_0$ of (\ref{Hypo}). Under the alternative hypothesis, we considered one change-point $\tau \in \{1, \cdots, n-1\}$ such as $\mu_i=0$ for $i \le \tau$ and $\mu_i=\mu$ for $\tau+1 \le i \le n$. The non-zero mean vector $\mu$ had $[p^{0.7}]$ non-zero components which were uniformly and randomly drawn from $p$ coordinates $\{1, \cdots, p\}$. Here, $[a]$ denotes the integer part of $a$. The magnitude of non-zero entry of $\mu$ was controlled by a constant $\delta$ multiplied by a random sign. 
The nominal significance level was chosen to be $0.05$. All the simulation results were obtained based on 1000 replications.

We also considered {two competitors. One is the E-div test proposed by Matteson and James (2014) and the other is the CQ test proposed by Chen and Qin (2010). 
Both testing procedures assume independence of $\{X_i\}_{i=1}^n$. 
The CQ test was originally designed for the two-sample problem, requiring the change point to be known. 
To implement the CQ test, we used the true change point to divide the sequence into two samples under the alternative hypothesis. Under the null hypothesis,  we obtained the two samples by adopting the same time point used for the alternative.

Table \ref{case1} demonstrates empirical sizes and powers of three tests with Gaussian $\epsilon_i$ in (\ref{data_g}). To obtain power, we chose the location of the change point $\tau=0.4 n$ and magnitude $\delta=0.3$. Under temporal independence ($M=0$), sizes of all three tests were well controlled around the nominal significance level $0.05$. Under dependenc ($M=1, 2$), the CQ and E-div tests suffered severe size distortion. Unlike those two tests, the proposed test still had sizes well controlled around the nominal significance level $0.05$. Due to severe size distortion of the CQ and E-div tests, it is not relevant to compare the power of three tests under dependence. We thus only conducted power comparison when $M=0$. Empirical powers of three tests increased as $n$ and $p$ increased. The reason the CQ test had the best power among three is that it utilized the information of location of the change point. In real application, such information is unavailable. The proposed test always enjoyed greater powers than the E-div test with respect to different $n$ and $p$.

\begin{figure}[t!]
\begin{center}
\includegraphics[width=0.45\textwidth,height=0.45\textwidth]{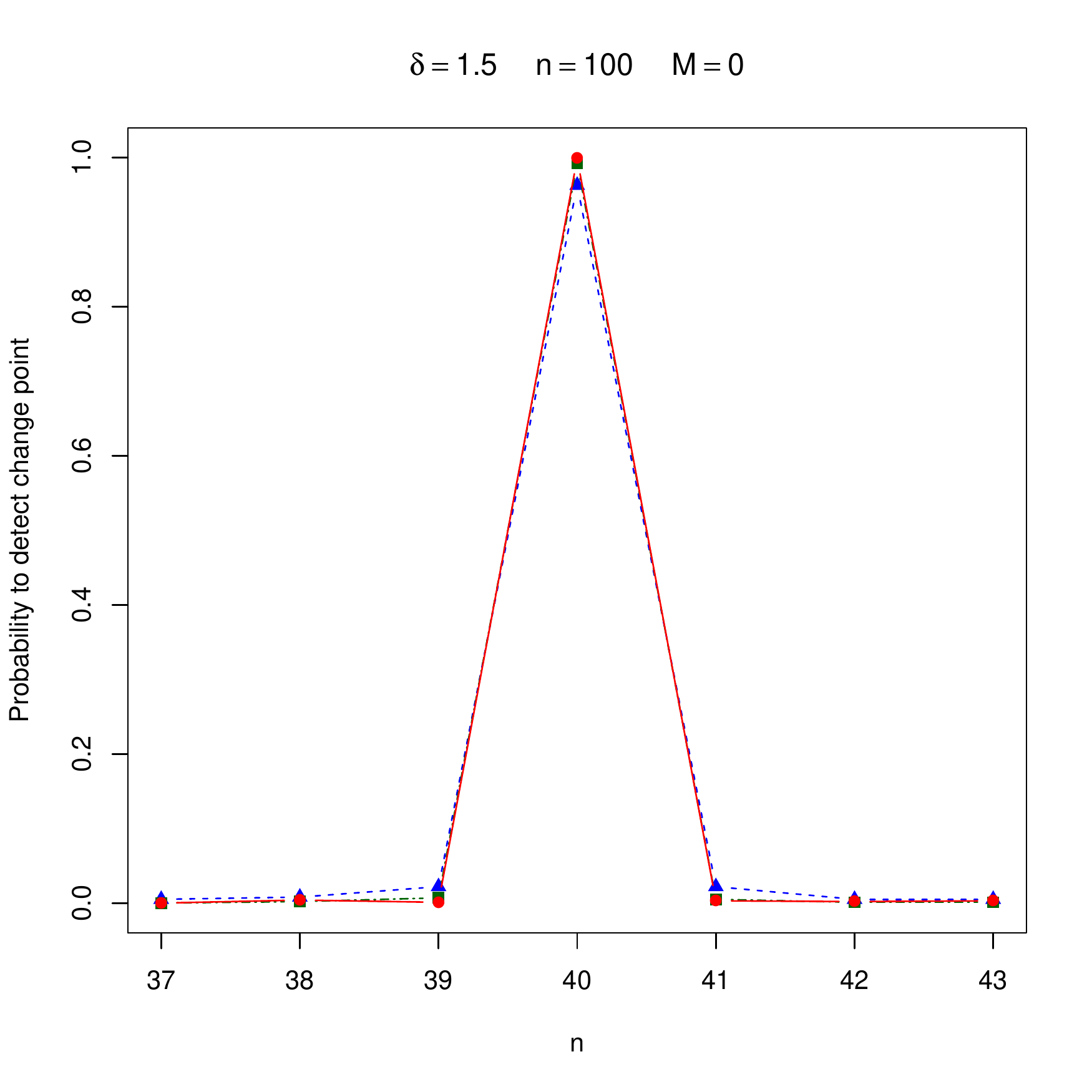}
\includegraphics[width=0.45\textwidth,height=0.45\textwidth]{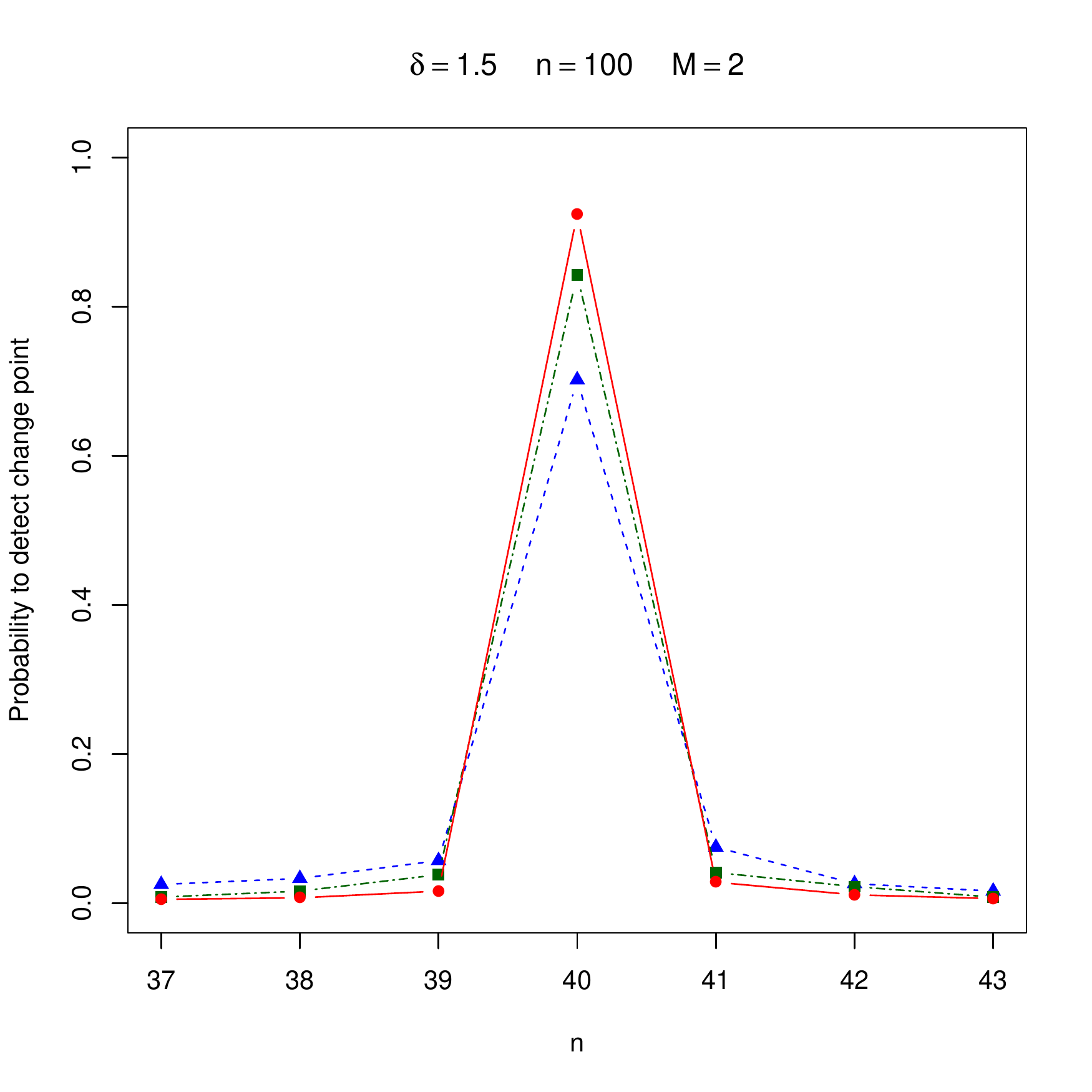}\\
\includegraphics[width=0.45\textwidth,height=0.45\textwidth]{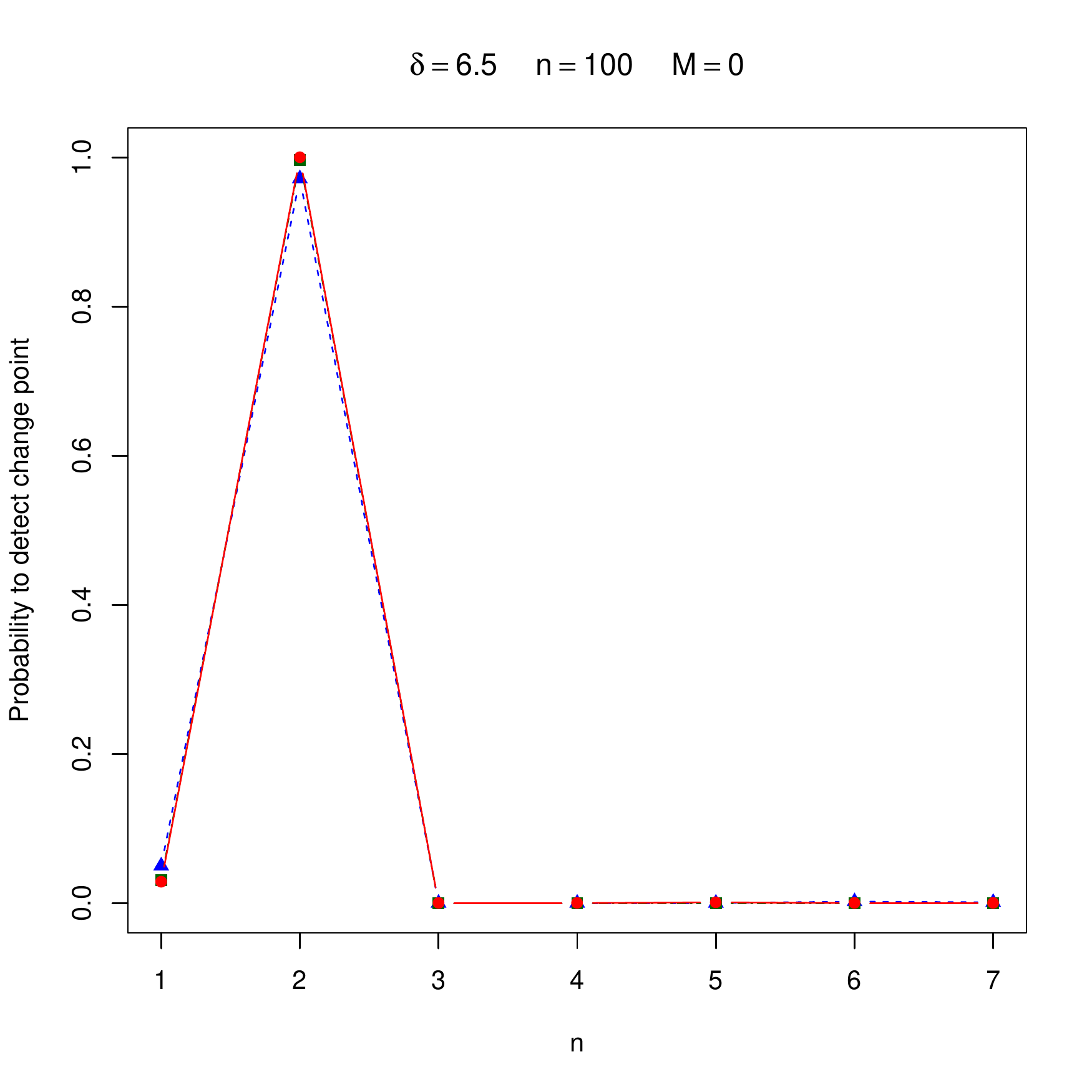}
\includegraphics[width=0.45\textwidth,height=0.45\textwidth]{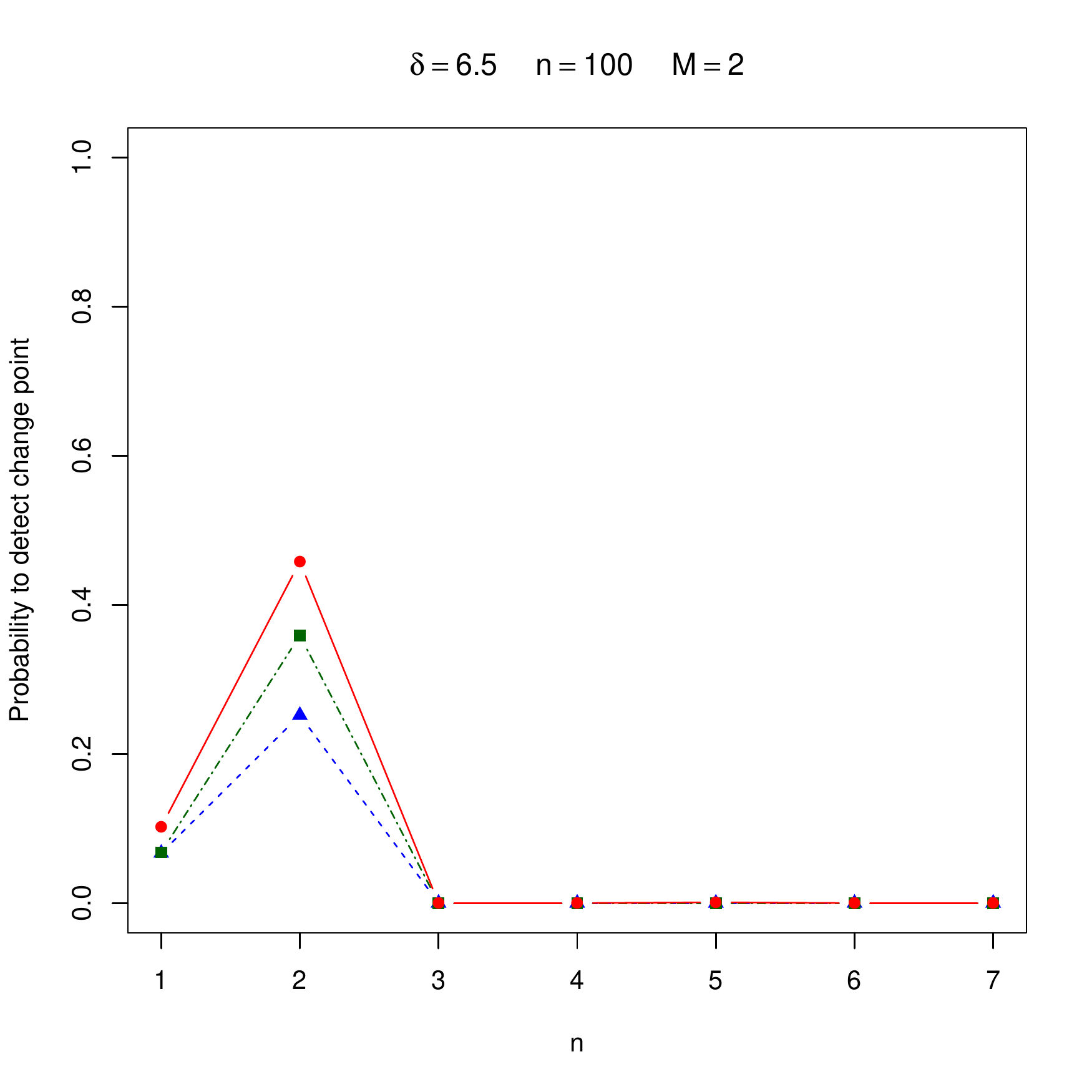}
\caption{The probability of detecting a change point when $p=100$ (trangles), $300$ (squares) and $600$ (circles). Upper panel: the change point is at $40$. Lower panel: the change point is at $2$. }
\label{fig3}
\end{center}
\end{figure}

Under spatial and temporal dependence, we also studied the power of the proposed test subject to different combinations of sample size $n$, dimension $p$ and location of the change point. The results are included in the supplementary material. 
The supplementary material also includes the results of the proposed test when $\epsilon_i$ follows $t$-distribution. 
The patterns of sizes and powers were quite similar to those when $\epsilon_i$ follows Gaussian distribution, showing the nonparametric property of the proposed test. 

\subsection*{4.2 \bf Empirical performance of the estimating procedure}

The second part of the simulation studies aims to investigate the empirical performance of the change point estimator ${\tau}_e$ in (\ref{identification}). We first considered the situation with one change-point $\tau \in \{1, \cdots, n-1\}$ such as $\mu_i=0$ for $i \le \tau$ and $\mu_i=\mu$ for $\tau+1 \le i \le n-1$. The non-zero mean vector $\mu$ had $[p^{0.7}]$ non-zero components, which were uniformly and randomly drawn from $p$ coordinates $\{1, \cdots, p\}$. The magnitude of non-zero entry of $\mu$ was controlled by a constant $\delta$ multiplied by a random sign.
Figure \ref{fig3} demonstrates the proportion of the $1000$ iterations detecting the change point that was located at the time point $40$ and $2$, respectively. First, the probability of detecting the change point increased as dimension $p$ increased.
Second, comparing the right panel with the left panel, the probability of detecting the change point became lower as dependence increased from $M=0$ to $2$. Finally, comparing the lower panel with the upper panel,  the stronger signal strength was needed when the change point was at $2$, in order to retain the similar detection probability when the change point was located at $40$. Our empirical results are consistent with the theoretical results in Theorem 4.

We also compared the proposed change-point estimator with the one proposed in Bai (2010).
Since the estimating procedure in Bai (2010) assumes that the change point exists {\it a priori}, we implemented both methods without conducting hypothesis testing. 
Figure \ref{f3-2-2} illustrates that the change-point estimator in Bai (2010) failed to identify the change point at $2$ under temporal dependence ($M=2$). Unlike the change-point estimator in Bai (2010), the proposed change-point estimator performed well under both temporal independence and dependence. 

\begin{figure}[t!]
\begin{center}
\includegraphics[width=0.45\textwidth,height=0.45\textwidth]{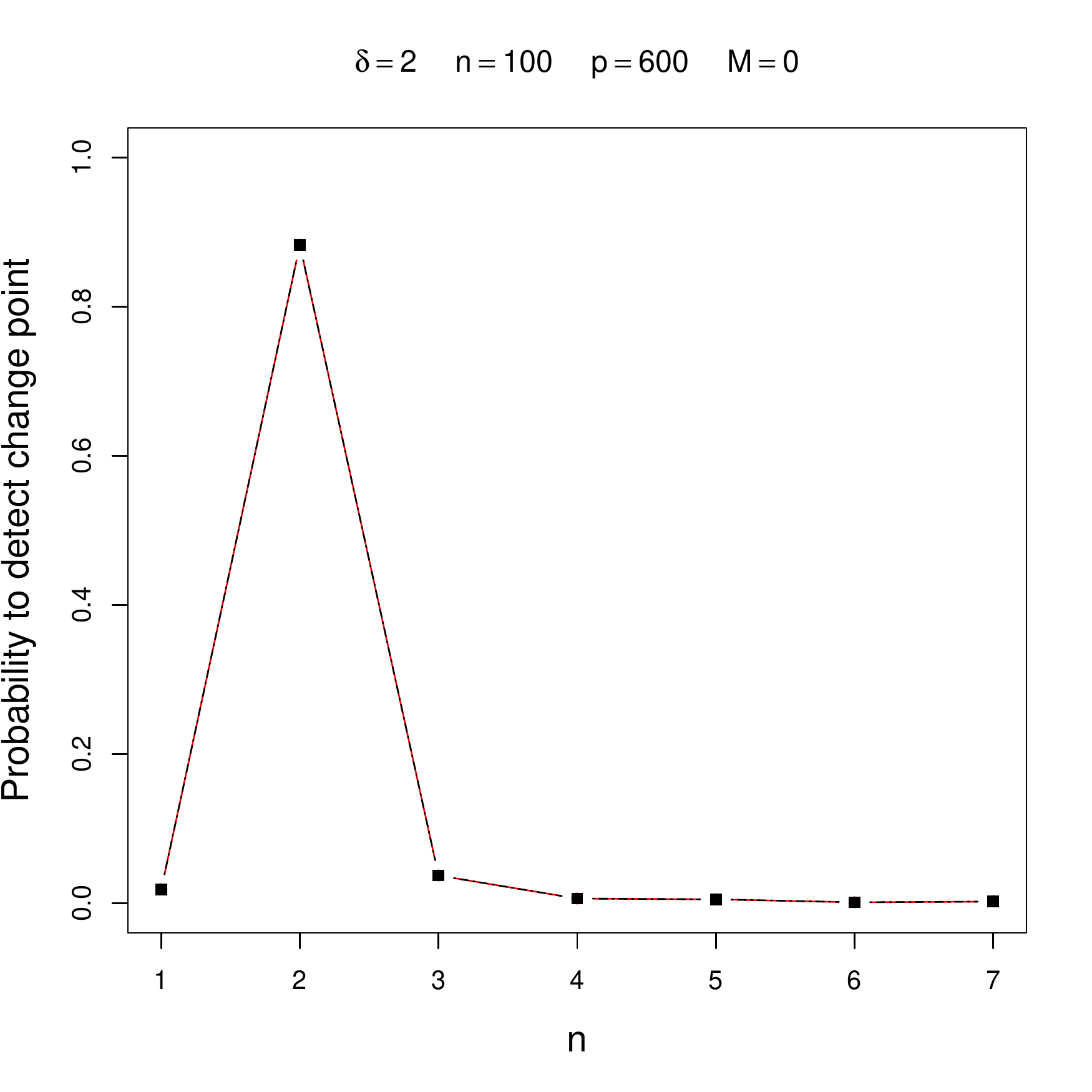}
\includegraphics[width=0.45\textwidth,height=0.45\textwidth]{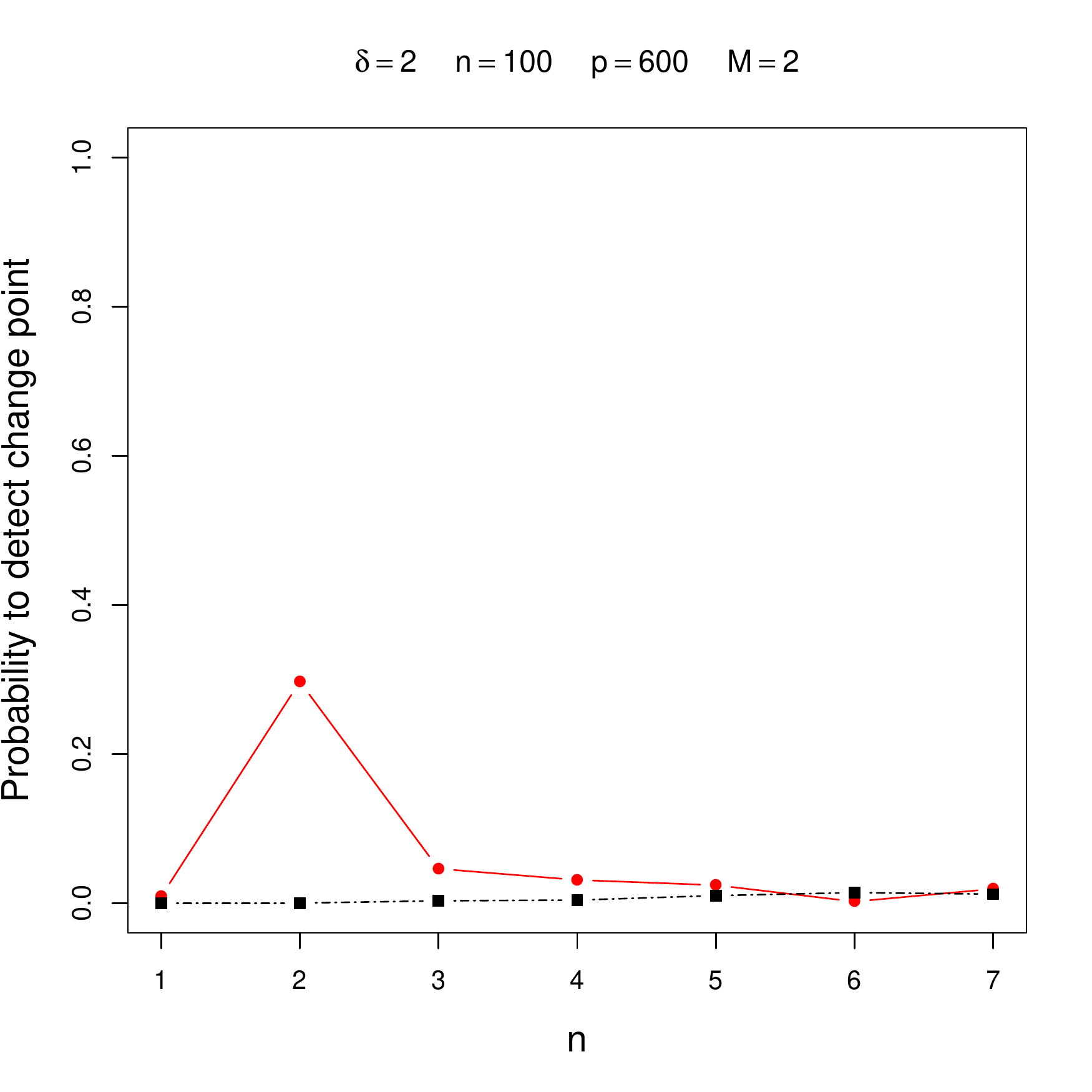}\\
\caption{The probability of detecting a change point at 2 by the proposed estimator (circles) and the estimator of Bai (2010) (squares).  Left panel: data are temporally independent with $M=0$. Right panel: data are temporally dependent with $M=2$. }
\label{f3-2-2}
\end{center}
\end{figure}

\begin{table}[tbh!]
\tabcolsep 3.8pt
\begin{center}
\caption{The performance of the proposed binary segmentation and E-div method for estimating multiple change points with Gaussian $\epsilon_i$ in (\ref{data_g}). The average FP, FN, TP and corresponding standard deviations were obtained based on $1000$ replications. }
\label{case2}
\begin{tabular}{cccccccccc}
 &&\multicolumn{2}{c}{$p=200$} &&\multicolumn{2}{c}{$600$} &&\multicolumn{2}{c}{$1000$}\\[1mm]
$M$ & & New & E-div & & New & E-div & & New& E-div  \\
\multicolumn{10}{c}{$(\delta_1,\delta_2)=(0.5,0.5)$}                                                                                                                             \\
    & FP & $1.068_{0.703}$         & $0.816_{0.577}$           &  & $1.074_{0.762}$         & $0.880_{0.645}$           &  & $1.050_{0.767}$         & $0.879_{0.635}$           \\
0   & FN & $2.785_{0.421}$         & $2.832_{0.374}$           &  & $2.717_{0.462}$         & $2.762_{0.435}$           &  & $2.688_{0.485}$         & $2.693_{0.472}$           \\
    & TP & $0.215_{0.421}$         & $0.168_{0.374}$           &  & $0.283_{0.462}$         & $0.238_{0.435}$           &  & $0.321_{0.485}$         & $0.307_{0.472}$           \\
[0.25cm]
    & FP & $0.258_{0.498}$         & $19.974_{2.702}$          &  & $0.316_{0.530}$         & $22.187_{1.293}$          &  & $0.454_{0.573}$         & $22.120_{1.261}$          \\
2   & FN & $2.982_{0.133}$         & $2.474_{0.660}$           &  & $2.982_{0.133}$         & $2.431_{0.674}$           &  & $2.961_{0.194}$         & $2.394_{0.764}$           \\
    & TP & $0.018_{0.133}$         & $0.526_{0.660}$           &  & $0.018_{0.133}$         & $0.569_{0.674}$           &  & $0.039_{0.194}$         & $0.606_{0.704}$           \\
\multicolumn{10}{c}{$(\delta_1, \delta_2)=(1.5, 1.5)$}\\
     &FP  & $0.210_{0.454}$ & $0.149_{0.396}$ &  & $0.078_{0.279}$ & $0.083_{0.300}$& & $0.033_{0.190}$& $0.057_{0.240}$ \\
 $0$ & FN  & $0.153_{0.390}$ & $0.094_{0.318}$&  & $0.047_{0.212}$ & $0.022_{0.147}$&  & $0.022_{0.147}$& $0.006_{0.077}$ \\
  &   TP  & $2.847_{0.390}$ & $2.906_{0.318}$ &  & $2.953_{0.212}$ & $2.978_{0.147}$& & $2.977_{0.146}$& $2.994_{0.077}$ \\[0.25cm]
     &FP  & $0.502_{0.612}$ & $18.619_{2.864}$ &  & $0.274_{0.495}$ & $20.336_{1.434}$& & $0.206_{0.436}$& $20.187_{1.360}$ \\
 $2$ & FN  & $2.162_{0.648}$ & $1.094_{0.877}$&  & $1.889_{0.641}$ & $0.599_{0.735}$&  & $1.706_{0.751}$& $0.421_{0.647}$ \\
  &   TP  & $0.838_{0.648}$ & $1.906_{0.877}$ &  & $1.111_{0.641}$ & $2.401_{0.735}$& & $1.294_{0.751}$& $2.579_{0.647}$ \\
\end{tabular}
\end{center}
\end{table}

The last part of the simulation studies is to demonstrate the performance of the proposed binary segmentation method for multiple change-point detection. We chose $n=150$ and considered three change points at $15$, $75$ and $105$, respectively. 
In particular, for $1\le i \le 15$, $\mu_i=0$. For $16 \le i \le 75$, the non-zero entry of $\mu_i$ was controlled by a constant $\delta_1$. For $76 \le i \le 105$, $\mu_i=0$ and for $106 \le i \le 150$, the non-zero entry of $\mu_i$ was controlled by another constant $\delta_2$.  We compared our method with E-div method 
in terms of false positives (FP), false negatives (FN), and true positives (TP). The FP is the number of time points that are wrongly estimated as change points. The FN is the number of change points that are wrongly treated as time points without change. 
And TP is the total number of identified change points. A procedure is better if it has smaller FP and FN, but TP is close to $3$ which is the total number of change points based on our design. Table \ref{case2} demonstrates the performance of two methods based on $1000$ iterations when estimating the three change points. Under temporal independence ($M=0$), the two methods had similar performance with both FP and FN decreased but TP increased as $p$ and/or $(\delta_1, \delta_2)$ increased.
On the other hand, under temporal dependence ($M=2$), the E-div procedure suffered severe FP for all cases although it had larger TP. Different from the E-div procedure, the proposed method always had the FP and FN under control. Most importantly, similar to the case of $M=0$, it enjoyed smaller FP and FN but larger TP as $p$ and/or $(\delta_1, \delta_2)$ increased.




\setcounter{section}{5} \setcounter{equation}{0}
\section*{\large 5. \bf Application}

Southwest University, China conducted an fMRI experiment to exam the differences in brain activation between overweight and normal weight subjects when performing a body image self-reflection task. In the task, participants were instructed to view several fat and thin body images closely, and vividly imagine that someone was comparing her body to the body in the picture. The experiment comprised six blocks of the fat body condition and six blocks of the thin body condition, and each block consisted of seven images. During the experiment, the brain of each participant was scanned every 2 seconds and total 280 images were taken, and each of image consisted of 131,072 voxels. Hence, for each subject, the high-dimensional time course data have $p=131,072$ and $n=280$. The recorded fMRI data are publicly available at https://openfmri.org/dataset/ds000213/.

\begin{figure}[t!]
\begin{minipage}[t]{0.45\linewidth}
\centering
\subfloat[Normal weight, thin-body image]{\label{fig1a}%
\includegraphics[width=2.5in]{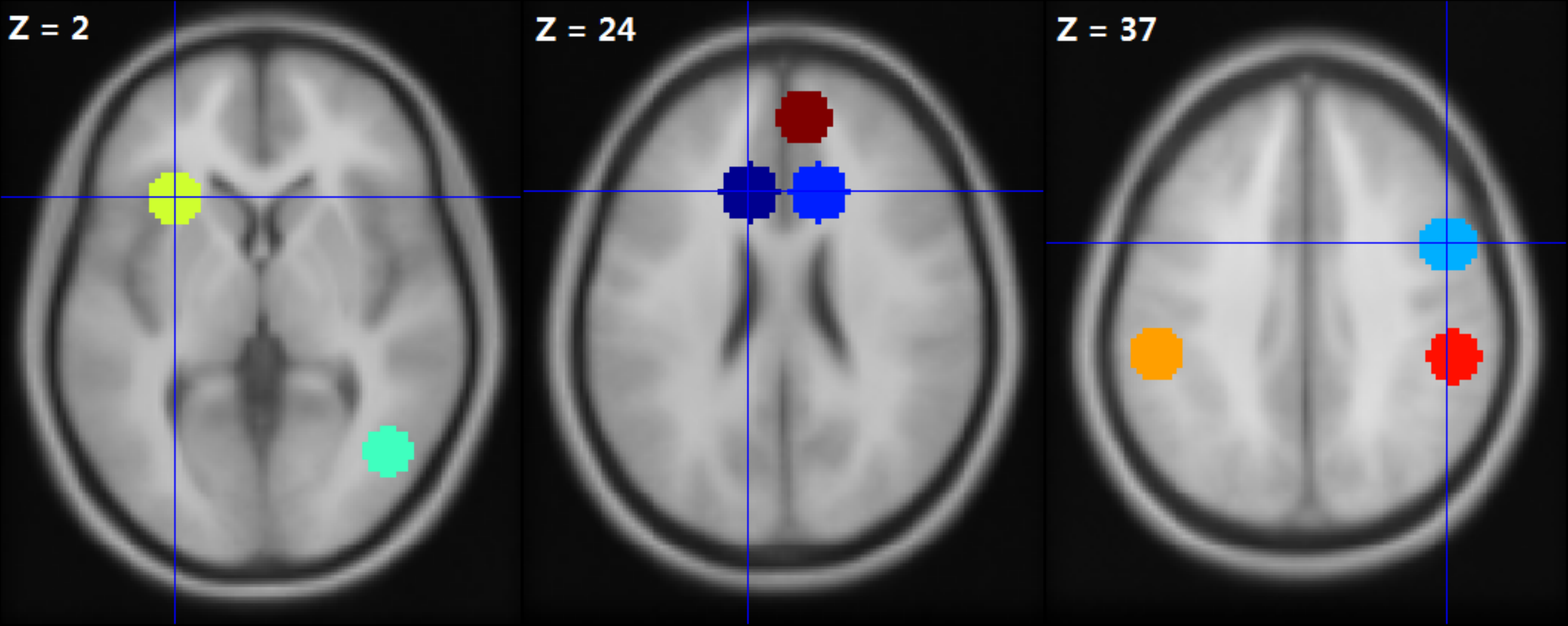}}\\%
\subfloat[Normal weight, fat-body image]{\label{fig1c}%
\includegraphics[width=2.5in]{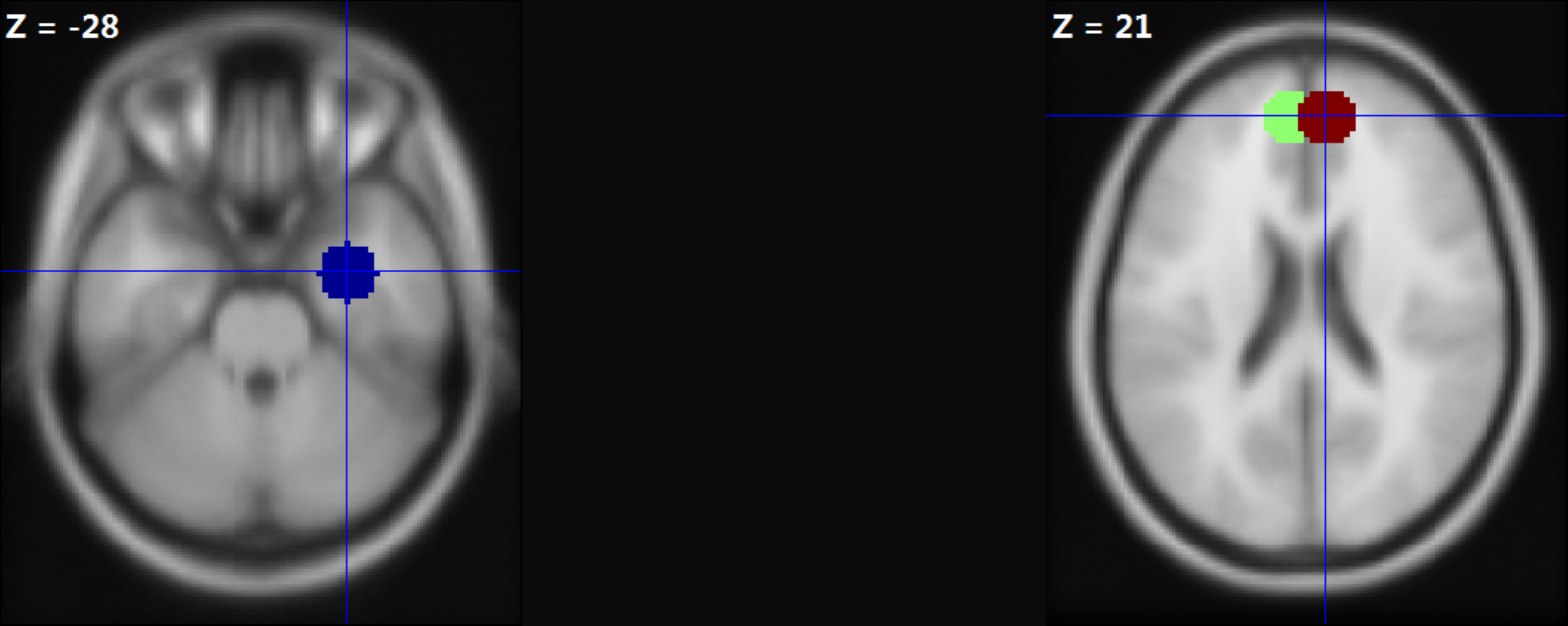}}
\end{minipage}
\begin{minipage}[t]{0.5\linewidth}
\centering
\subfloat[Overweight, thin-body image]{\label{fig1b}%
\includegraphics[width=2.5in]{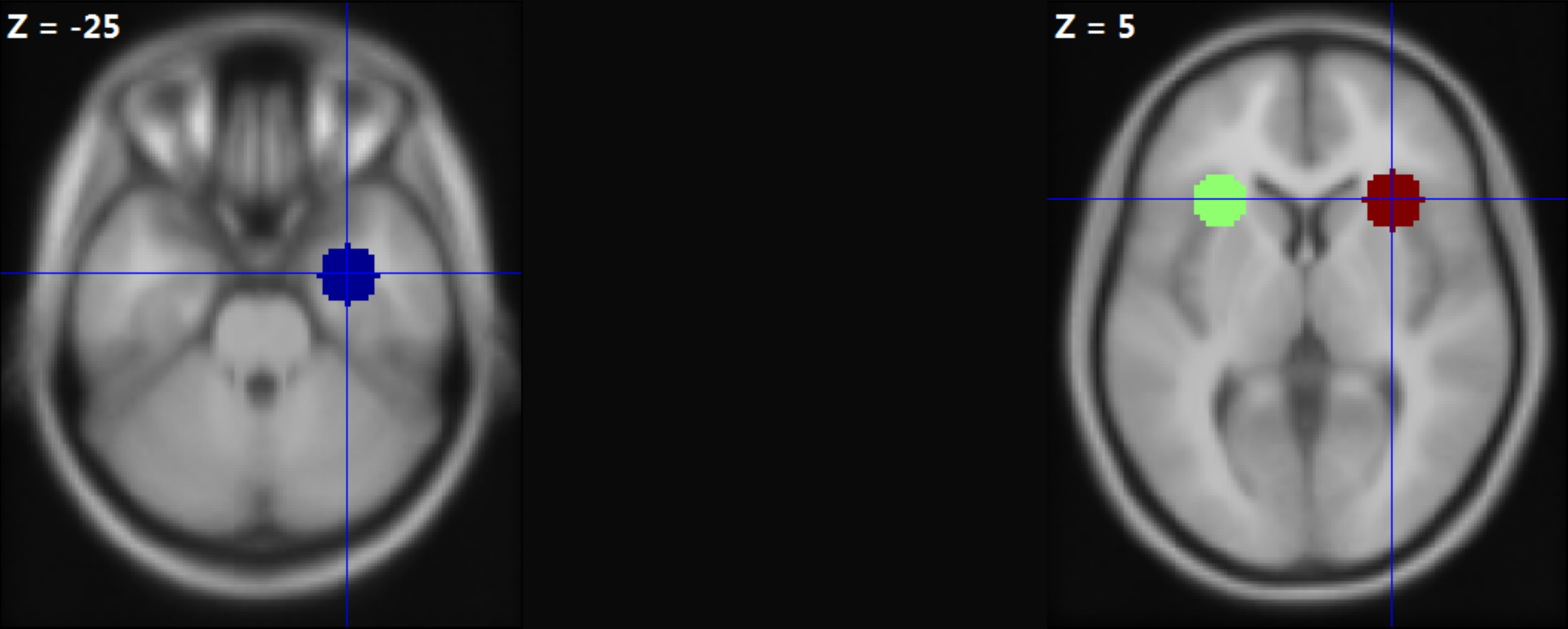}}\\%
\subfloat[Overweight, fat-body image]{\label{fig1d}%
\includegraphics[width=2.5in]{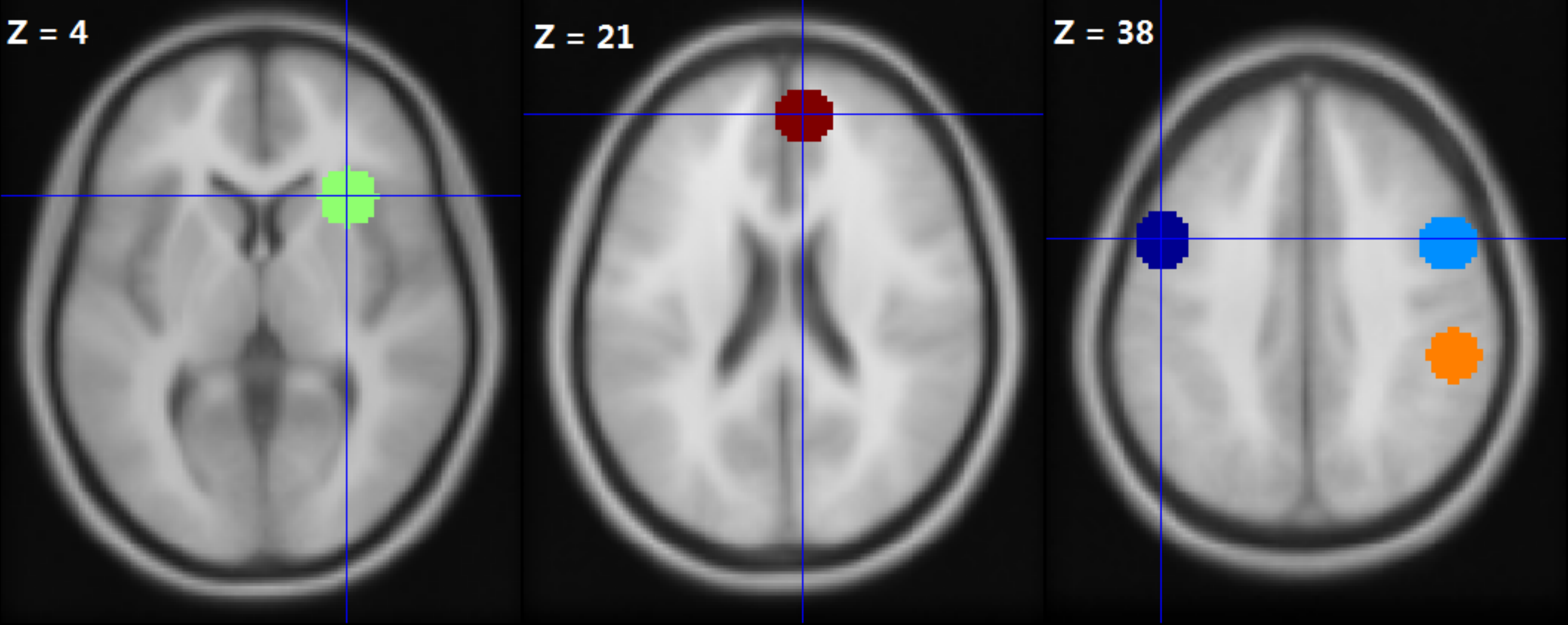}}%
\end{minipage}
\caption{Activation map for ROIs. In panel (a): left insula (yellow); right EBA (cyan); left ACC (darkblue); right ACC (blue); right MPFC (darkred); left IPL (orange); right IPL (red); right DLPFC (deepskyblue). In panel (b): right amygdala (darkblue); left MPFC (lightgreen); right MPFC (darkred). In panel (c): right FBA (darkblue); left insula (lightgreen); right insula (darkred). In panel (d): right insula (lightgreen); right MPFC (darkred); left DLPFC (darkblue);  right DLPFC (deepskyblue); right IPL (orange).}
\label{fig6}
\end{figure}

We randomly picked normal weight subject 7 and overweight subject 1 from the fMRI data. Based on Gao et al.(2016), we used the MNI coordinates to partition all voxels into 16 distinct regions of interest (ROIs) which play different functions. For example, previous studies found that inferior parietal lobule (IPL), extrastriate body area (EBA, lateral occipitotemporal cortex) and fusiformbody area (FBA) were related to perceptive processing of body image; Dorsolateral prefrontal cortex (DLPFC) and amygdala were related to affective processing of body image and can be activated when viewing body pictures with negative emotional valence; Medial prefrontal cortex (MPFC) was related with self-reflection; and ACC and insula were related to body dissatisfaction (Wagner et al., 2003; Uher et al., 2005; Kurosaki et al., 2006; Friederich et al., 2007; Miyake et al., 2010; Friederich et al., 2010; Yang et al., 2014).

Although the time points when the different types of images were applied to each subject are known, we treated the data as if such information were not available in advance. 
We applied the proposed change-point detection method to 16 ROIs with the nominal significance level $\alpha=0.05$ and the $M$ dependence estimated by the elbow method. If an ROI was tested to encounter one or more change points, it was activated. Moreover, the type of image induced the change is known at each identified change point. Figure \ref{fig6} illustrates the physical locations of the ROIs activated (changes detected) by the thin-body images and fat-body images for the normal weight subject and the overweight subject. More precisely, 5 ROIs were activated for the overweight subject when viewing fat-body images, while only 3 ROIs were activated when viewing thin-body images. On the other hand, for the normal weight subject, 8 ROIs were activated when viewing thin-body images, while only 3 ROIs were activated when viewing fat-body images.


Our results indicate that the overweight subject showed a stronger visual processing of fat body images than thin body images, whereas the normal weight subject showed a stronger visual processing of thin body images than fat body images. Interestingly, we found that ACC was only activated for the normal weight women when viewing thin body images. Such a result was consistent with the findings in Friederich et al. (2007), that healthy women body dissatisfaction and self-ideal discrepancies can be greatly induced by exposure to attractive slim bodies of other women. This may be one reason that normal weight women are more motivated to watch their weight and keep in shape than the overweight women.

\section*{Reference}
\begin{description}
\item
Ayyala, D., Park, J. and Roy, A. (2017), ``Mean vector testing for high-dimensional dependent observations," \textit{Journal of Multivariate Analysis}, 153, 136-155.
\item
{Bai, J.} (2010), ``Common breaks in means and variances for panel data,ее \textit{Journal of Econometrics}, {157}, 78--92.

\item
Bai, Z. D. and Saranadasa, H. (1996), ``Effect of high dimension: By an example of a two sample problem," \textit{Statistica Sinica}, 6, 311-329.
\item
Carrasco, M. and Chen, X. (2002), ``Mixing and moment properties of various GARCH and stochastic volatility models," \textit{Econometric Theory}, 18, 17-39.
\item
Chen, J. and Gupta, A. (1997), ``Testing and locating variance change-points with application to stock prices," \textit{Journal of the American Statistical Association}, 92, 739-747.
\item
Chen, S. X. and Qin, Y. (2010),`` A two-sample test for high-dimensional data with applications to gene-set testing," \textit{The Annals of Statistics}, 38, 808-835.
\item
Chen, H. and Zhang, N. (2015), ``Graph-based change-point detection," \textit{The Annals of Statistics}, 43, 139-176.
\item
Davis, R. A., Lee, T. and Rodriguez- Yam, G (2006), ``Structural break estimation for non-stationary time series," \textit{Journal of the American Statistical Association}, 101, 223-239.
\item
--------(2008), ``Break detection for a class of nonlinear time series models," \textit{Journal of Time Series Analysis}, 29, 834-867.
\item
Desobry, F., Davy, M. and Doncarli, C. (2005), ``An online kernel change detection algorithm," \textit{Signal Processing, IEEE Transaction on}, 53, 2961-2974.
\item
Friederich, H. C., Brooks, S., Uher, R., Campbell, I. C., Giampietro, V., Brammer, M., Williams, S.C.R., Herzog, W., and  Treasure, J. (2010), ``Neural correlates of body dissatisfaction in anorexia nervosa," \textit{Neuropsychologia}, 48, 2878-2885.
\item
Friederich, H. C., Uher, R., Brooks, S., Giampietro, V., Brammer, M., Williams, S. C., Herzog, W.,Treasure, J., and Campbell, I. (2007), ``I'm not as slim as that girl: neural bases of body shape self-comparison to media images," \textit{Neuroimage}, 37, 674-681.
\item
Gao, X., Deng, X., Wen, X., She, Y., Vinke, P. and Chen, H. (2016), ``My body looks like that girl's: body mass index modulates brain activity during body images self-reflection among young women," \textit{PLoS ONE}, 11, e0164450.
\item
Harchaoui, Z., Moulines, E. and Bach, F. (2009), ``Kernel change-point analysis," \textit{Advances in Neural Information Processing Systems}, 609-616.
\item
Incl$\acute{a}$n, C. and Tiao, G. (1994), ``Use of sums of squares for retrospective detection of changes of variance," \textit{Journal of the American Statistical Association}, 89, 913-923.
\item
James, B., James, K. L. and Siegmund, D. (1992), ``Asymptotic approximations for likelihood ratio tests and confidence regions for a change-point in the mean of a multivariate normal distribution," \textit{Statistica Sinica}, 2, 69-90.
\item
Kokoszka, P. and Leipus, R. (2000), ``Change-point estimation in ARCH models,"  \textit{Bernoulli}, 6, 513-539.
\item
Kurosaki, M., Shirao, N., Yamashita, H., Okamoto, Y. and Yamawaki, S. (2006), ``Distorted images of one's own body activates the prefrontal cortex and limbic/paralimbic system in young women: a functional magnetic resonance imaging study," \textit{Biological Psychiatry}, 59, 380-386.
\item
Lavielle, M. and Moulines, E. (2000), ``Least-squares estimation of an unknown number of shifts in a time series," \textit{Journal of Time Series Analysis}, 21, 33-59.
\item
{Li, J. and Chen, S. X.} (2012), ``Two sample tests for high dimensional covariance matrices,ее \textit{The Annals of Statistics}, {40}, 908--940.
\item
{Liu, W. and Shao, Q.} (2013), ``A Cramer moderate deviation theorem for Hotelling $T^2$-statistic with applications to global tests,ее \textit{The Annals of Statistics}, {41}, 296--322. 

\item
Matteson, D. and James, N. A. (2014), ``A nonparametric approach for multiple change point analysis of multivariate data," \textit{Journal of the American Statistical Association}, 109, 334-345.
\item
Miyake, Y., Okamoto, Y., Onoda, K., Kurosaki, M., Shirao, N. and Yamawaki, S. (2010), ``Brain activation during the perception of distorted body images in eating disorders," \textit{Psychiatry Research: Neuroimaging}, 181, 183-192.
\item
{Okamoto, J., Stewart, N. and Li, J.} (2018). HDcpDetect: detect change points in means of high dimensional data. R Package Version 0.1.0. (Available from https://cran.r-project.org/web/packages/HDcpDetect/.)
\item
{Olshen, A. and Venkatraman, E.} (2004), ``Circular binary segmentation for the analysis of array-based DNA copy number data,ее \textit{Biostatistics}, {5}, 557--572.

\item
Ombao, H. C., Raz, J. A., von Sachs, R. and Molow, B. A. (2001), ``Automatic statistical analysis of bivariate nonstationary time series," \textit{Journal of The American Statistical Association}, 96, 543-560.
\item
Sen, A. K. and Srivastava, M. S. (1975), ``On tests for detecting change in mean," \textit{The Annals of Statistics}, 3, 98-108.
\item
Shao, X. and Zhang, X. (2010), ``Testing for change points in time series," \textit{Journal of the American Statistical Association}, 105, 1228-1240.
\item
Siegmund, D., Yakir, B. and Zhang, N. R. (2011). ``Detecting simultaneous variant intervals in aligned sequences," \textit{The Annals of Applied Statistics}, 5, 645-668.
\item
Srivastava, M.S. and Worsley, K. J. (1986). ``Likelihood ratio tests for a change in the multivariate normal mean," \textit{Journal of the American Statistical Association}, 81, 199-204.
\item
Uher, R., Murphy, T., Friederich, H. C., Dalgleish, T., Brammer, M. J. and Giampietro, V.(2005), ``Functional neuroanatomy of body shape perception in healthy and eating-disordered women," \textit{Biological Psychiatry}, 58, 990-997.
\item
Vostrikova, L. (1981), ``Detection of disorder in multidimensional random processes," \textit{Soviet Mathematics Doklady}, 24, 55-59.
\item
Wagner, A., Ruf, M., Braus, D. F. and Schmidt, M. H. (2003), ``Neuronal activity changes and body image distortion in anorexia nervosa," \textit{Neuroreport}, 14, 2193-2197.
\item
Yang, J., Dedovic, K., Guan, L., Chen, Y. and Qi, M. (2014), ``Self-esteem modulates dorsal medial prefrontal cortical response to self-positivity bias in implicit self-relevant processing," \textit{Social Cognitive and Affective Neuroscience}, 9, 1814-1818.
\item
Zhang, N. R., Siegmund, D. O., Ji, H. and Li, J. Z. (2010), ``Detecting simultaneous changepoints in multiple sequences," \textit{Biometrika}, 97, 631-645.
\end{description}
\end{document}